\documentclass[journal]{IEEEtran}

\usepackage{cite}
\usepackage{caption}
\usepackage{amsmath,amssymb,amsfonts}
\usepackage{booktabs}
\usepackage{graphicx}
\usepackage{textcomp}
\usepackage{subfig}
\usepackage{algorithm}
\usepackage{algorithmic}
\usepackage{bm} 

\def\BibTeX{{\rm B\kern-.05em{\sc i\kern-.025em b}\kern-.08em
    T\kern-.1667em\lower.7ex\hbox{E}\kern-.125emX}}

\begin{document}


\title{
Enabling Large-Scale Channel Sounding for 6G: A Framework for Sparse Sampling and Multipath Component Extraction}

\author{
Yi Chen,~\IEEEmembership{Member,~IEEE}, Ming Li,~\IEEEmembership{Senior Member,~IEEE}, Chong Han,~\IEEEmembership{Senior Member,~IEEE}
\thanks{Yi Chen and Ming Li are with the school of information and communication engineering, Dalian University of Technology, Dalian, China (e-mail: chenyi@dlut.edu.cn, and mli@dlut.edu.cn).
\par Chong Han is with Terahertz Wireless Communication (TWC) center, Shanghai Jiao Tong University, Shanghai, China (e-mail: chong.han@sjtu.edu.cn).
}
}
\maketitle

\begin{abstract}
Realizing the 6G vision of artificial intelligence (AI) and integrated sensing and communication (ISAC) critically requires large-scale real-world channel datasets for channel modeling and data-driven AI models. However, traditional frequency-domain channel sounding methods suffer from low efficiency due to a prohibitive number of frequency points to avoid delay ambiguity. This paper proposes a novel channel sounding framework involving sparse nonuniform sampling along with a likelihood-rectified space-alternating generalized expectation-maximization (LR-SAGE) algorithm for multipath component extraction. This framework enables the acquisition of channel datasets that are tens or even hundreds of times larger within the same channel measurement duration, thereby providing the massive data required to harness the full potential of AI scaling laws. Specifically, we propose a Parabolic Frequency Sampling (PFS) strategy that non-uniformly distributes frequency points, effectively eliminating delay ambiguity while reducing sampling overhead by orders of magnitude. To efficiently extract multipath components (MPCs) from the channel data measured by PFS, we develop a LR-SAGE algorithm, rectifying the likelihood distortion caused by nonuniform sampling and molecular absorption effect. Simulation results and experimental validation at 280--300~GHz confirm that the proposed PFS and LR-SAGE algorithm not only achieve 50$\times$ faster measurement, a 98\% reduction in data volume and a 99.96\% reduction in post-processing computational complexity, but also successfully captures MPCs and channel characteristics consistent with traditional exhaustive measurements, demonstrating its potential as a fundamental enabler for constructing the massive ISAC datasets required by AI-native 6G systems.
\end{abstract}

\begin{IEEEkeywords}
Channel sounding, channel modeling, mmWave and THz communications, 6G, artificial intelligence, integrated sensing and communication
\end{IEEEkeywords}
\section{Introduction}
    \par The vision for 6G outlined by the IMT-2030 framework positions the convergence of Artificial Intelligence (AI) and Integrated Sensing and Communication (ISAC) as a cornerstone of future wireless systems, fundamentally altering the role of the wireless channel~\cite{wcx}. Unlike conventional communication-centric models that abstract the physical world into simplified statistics, ISAC must exploit detailed channel information for environmental perception. This new paradigm, coupled with the data-hungry nature of AI models designed to unlock unprecedented performance, creates an urgent demand for adequate data that mirrors real-world propagation complexities~\cite{AI-ISAC}. Indeed, evidence suggests that synthetic data from statistical channel models fail to capture these complexities, leading to flawed performance evaluations and hindering AI's potential in 6G~\cite{critical}. Consequently, the development of large-scale, real-world ISAC channel datasets has become a critical prerequisite for advancing 6G research~\cite{ywf}.
    \par The acquisition of large-scale channel datasets is currently hampered by the inherent inefficiency of existing wideband channel sounding methodology. To cover bandwidths of tens of gigahertz in mmWave/THz bands and maintain precise phase coherence, vector-network-analyzer (VNA)-based frequency-domain channel sounding is generally adopted, which employs uniform frequency sampling~\cite{thz-cst,cxs-rof,ju-office,ju-factory,lyj,gk}. This approach is governed by the Nyquist criterion, which dictates a number of frequency samples proportional to both channel bandwidth and the maximum path length to avoid ``Delay Ambiguity'', a phenomenon where multipath components (MPCs) with delays exceeding the maximum unambiguous delay range (UDR) appear as spurious paths at shorter delays, thereby corrupting the channel impulse response~\cite{zjh,na-d2d,na-cellular}. 
    Consequently, for wideband and long-range systems, this necessitates an enormous number of frequency points, rendering measurement durations impractically long, often spanning several hours for a single measurement location~\cite{cy-200G,na-icc,kurner}. This challenge is further exacerbated by the complex propagation physics at mmWave/THz frequencies, where molecular absorption superimposes frequency-selective attenuation onto the MPCs, and distorts channel impulses~\cite{multi-ray,molisch}. 
    Therefore, a highly efficient channel sounding framework that can significantly accelerate channel measurement procedure and accurately extract MPCs from channels distorted by molecular absorption is essential to enable large-scale channel datasets needed to drive 6G innovation.
 
 \par In the literature, sparse sampling techniques, which can reduce sampling overhead in spatial and frequency domains, have been extensively studied in the areas of antenna array theory, radar imaging and sounding. They are broadly categorized into two main paradigms: random sampling based on Compressive Sensing (CS) theory~\cite{bcs,sbi}, and deterministic nonuniform sampling~\cite{unequal,Superresolution}. Random sparse sampling leverages the inherent sparsity of the channel in the delay domain to enable reconstruction from a reduced set of measurements. This paradigm has found successful applications in various fields, such as radar imaging, and ground-penetrating Radar~\cite{cs-sfcw,cs-thz}. However, its application in channel sounding is challenged by designing a random sampling matrix that provably satisfies the restricted isometry property and maintains low coherence for guaranteed sparse recovery. In contrast, deterministic nonuniform sampling provides a more structured methodology originating in the spatial domain for antenna arrays to achieve super-resolution direction-of-arrival (DoA) estimation~\cite{nested,coprime}. By arranging a small number of antenna elements in a nonuniform pattern, such as a coprime or nested array, a much larger ``virtual'' aperture, can be synthesized, thereby increasing the degrees of freedom~\cite{coarray}. This method has been extended from the spatial domain to the frequency domain for radar applications~\cite{bcoprime,radar-coprime}. For instance, in stepped-frequency continuous-wave (SFCW) radar imaging, nonuniform sampling is designed to shape the ambiguity function to eliminate false targets and clutter~\cite{SFCW-SQRT,SFCW-PSF}. 
 
 \par
\begin{figure}[t]
    \centering
        \includegraphics[width=0.48\textwidth]{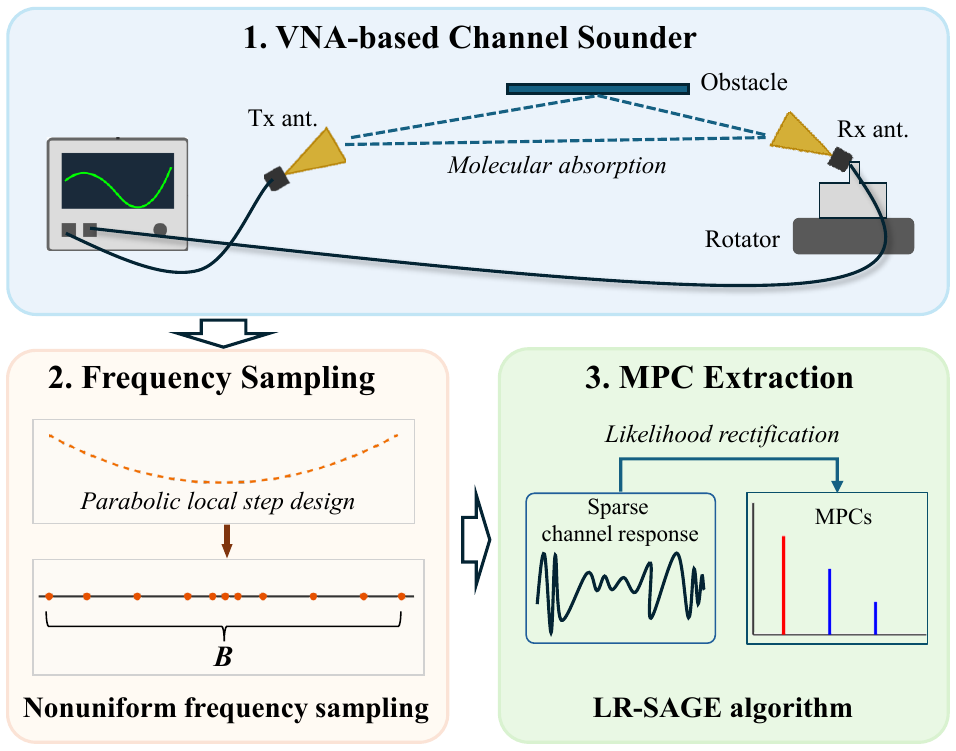}%
    \caption{Proposed sparse frequency sampling framework for channel sounding.}
    \label{fig:systemmodel}
    \vspace{-0.2cm} 
\end{figure}

 While the principles of sparse sampling are well-established in fields such as antenna array theory and radar sensing, their direct application to channel sounding is precluded by fundamental differences in the governing analytical models and performance objectives. Consequently, to the best of our knowledge, a holistic framework for sparse channel sounding has yet to be established and validated. This gap is further magnified in the mmWave and THz bands, where the physical phenomenon of molecular absorption introduces severe frequency-dependent distortion, an impairment that conventional MPC extraction algorithms fail to address. This paper bridges this critical gap by introducing a complete sparse channel sounding framework depicted in Fig.~\ref{fig:systemmodel}, comprising both a novel nonuniform frequency sampling scheme and a likelihood-rectified SAGE (LR-SAGE) algorithm engineered to jointly overcome these challenges. The core contributions of this paper are summarized as follows:

\begin{itemize}
    \item We present a performance analysis method for VNA-based channel sounding with SAGE-based MPC extraction in the mmWave and THz band by introducing a single-path likelihood function and considering general frequency sampling schemes and molecular absorption effect. Based on that, we analyze how the frequency sampling schemes and molecular absorption affect the likelihood function.
    \item We design a delay-ambiguity-free frequency sampling scheme for 6G channel sounding, i.e., parabolic frequency sampling (PFS). Compared with traditional uniform frequency sampling and nonuniform frequency sampling schemes with delay ambiguity, the designed scheme significantly reduces the number of frequency samples required, thus cutting down measurement time, data storage needs, and post-processing effort.
    \item We propose a LR-SAGE algorithm tailored for our designed PFS scheme to rectify the distorted likelihood function caused by nonuniform frequency sampling and molecular absorption. The proposed SAGE algorithm outperforms IDFT and the original SAGE algorithm in the mmWave and THz bands in terms of accurately extracting MPCs.
    \item We develop a THz channel sounding testbed with sparse sampling operating at 280–300 GHz to validate the proposed framework in real-world scenarios.  Experimental results corroborate that the proposed PFS and LR-SAGE algorithms achieve high-fidelity MPC extraction comparable to exhaustive uniform sampling, yet reduce the data volume and measurement time by over 98\% (i.e., a 50X acceleration). This demonstrates the practical viability of our sparse frequency sampling framework for constructing large-scale 6G channel datasets.
\end{itemize}

The remainder of this paper is organized as follows: In Section II, we formulate the ISAC channel sounding problem and describe the joint spatio-frequency channel model along with the classical SAGE algorithm. Section III provides an analysis of uniform, coprime, and nested frequency sampling schemes, quantifying their unambiguous delay ranges and the distortion due to molecular absorption. In Section IV, we introduce the delay-ambiguity-free parabolic frequency sampling scheme, including the guiding design principles and a theoretical guarantee of no ambiguity. Section V presents the LR-SAGE algorithm for MPC extraction, detailing the improvements over the conventional SAGE. Simulation results are given in Section VI to verify the performance gains of PFS and the LR-SAGE algorithm. Section VII shows experimental validation in a THz channel sounding scenario. Finally, Section VIII concludes the paper.
\section{System Model and Problem Formulation}
\label{sec:system_model}

In this section, we first present the mathematical model for the joint spatio-frequency channel sounding process, incorporating path-specific frequency- and delay-dependent gains caused by molecular absorption. Subsequently, we formulate the multipath parameter estimation problem and introduce the SAGE algorithm. Finally, we highlight the critical role of the single-path likelihood function in delay estimation of MPCs.
\subsection{SYSTEM MODEL AND PROBLEM FORMULATION}
\label{subsec:channel_model}

We consider a wideband multipath propagation environment characterized by $L$ distinct paths. Each path $l \in \{1, \dots, L\}$ is described by a set of parameters: its complex amplitude $\alpha_l \in \mathbb{C}$, time-of-arrival (delay) $\tau_l \in \mathbb{R}^+$, and its direction of arrival (DoA) specified by the azimuth angle $\phi_l \in [0, 2\pi)$ and elevation angle $\theta_l \in [0, \pi]$. The complete parameter set for the $l$-th path is denoted by the vector $\bm{\xi}_l = \{\alpha_l, \tau_l, \phi_l, \theta_l\}$. The channel is measured using a Vector Network Analyzer (VNA) connected to a directive horn antenna at the receiver as shown in Fig.~\ref{fig:systemmodel}. The measurement procedure for 6G ISAC channel typically involves two dimensions~\cite{DSS-SAGE}: 
\begin{enumerate}
    \item \textbf{Frequency Domain}: The VNA performs a frequency sweep over a set of $K$ discrete frequency points, $\mathcal{F} = \{f_1, f_2, \dots, f_K\}$. This set can be configured as either uniform frequency sampling (UFS) scheme, where $f_k = f_{\text{start}} + (k-1)\Delta f$, or a nonuniform sampling (NUS) scheme, such as coprime or nested sampling.
    \item \textbf{Spatial Domain}: After each frequency sweep, the horn antenna is mechanically rotated to a new pointing direction. The measurement is repeated for a set of $M$ distinct antenna pointing directions, $\bm{\Omega}_{\text{ant}} = \{\bm{\Omega}_1, \bm{\Omega}_2, \dots, \bm{\Omega}_M\}$, where each $\bm{\Omega}_m = (\phi^{\text{ant}}_m, \theta^{\text{ant}}_m)$ represents the azimuth and elevation of the antenna's main boresight.
\end{enumerate}
\par The channel frequency response (CFR) measured at the $k$-th frequency point $f_k$ and the $m$-th antenna pointing direction $\bm{\Omega}_m$ is the coherent superposition of the contributions from all $L$ paths, corrupted by additive noise. Thus, the measured signal $y(f_k, \bm{\Omega}_m)$ can be modeled as~\cite{multi-ray}:
\begin{align}
    y(f_k, \bm{\Omega}_m) =& \sum_{l=1}^{L} \alpha_l G_{\text{MA}}(f_k, \tau_l) G_{\text{ant}}(\phi_l, \theta_l | \bm{\Omega}_m) e^{-j 2\pi f_k \tau_l} \nonumber\\
    &+ n(f_k, \bm{\Omega}_m),
    \label{eq:scalar_model} 
\end{align} 
where $n(f_k, \bm{\Omega}_m)$ is the additive white Gaussian noise (AWGN), assumed to be a zero-mean, circularly symmetric complex Gaussian random variable with variance $\sigma^2$. The radiation pattern of the horn antenna is denoted by $G_{\text{ant}}(\phi, \theta | \bm{\Omega}_m)$, which describes the antenna's complex gain towards a direction $(\phi, \theta)$ when it is physically pointed towards $\bm{\Omega}_m$. Crucially, we introduce a path-specific gain term, $G_{\text{MA}}(f, \tau_l)$, which models the frequency-selective and delay-dependent attenuation caused by molecular absorption in the mmWave and THz bands.

For a more compact representation, we define a modified joint spatio-frequency steering vector $\bm{a}(\tau, \phi, \theta) \in \mathbb{C}^{KM \times 1}$. The $( (m-1)K + k )$-th element of this vector now incorporates the path gain:
\begin{equation}
    [\bm{a}(\tau, \phi, \theta)]_{k,m} = G_{\text{MA}}(f_k, \tau) G_{\text{ant}}(\phi, \theta | \bm{\Omega}_m) e^{-j 2\pi f_k \tau}.
\end{equation}
By stacking all $KM$ measurements into a single vector $\bm{y} \in \mathbb{C}^{KM \times 1}$, the overall channel model can be written in the familiar matrix-vector form:
\begin{equation}
    \bm{y} = \sum_{l=1}^{L} \alpha_l \bm{a}(\tau_l, \phi_l, \theta_l) + \bm{n} = \mathbf{A}(\bm{\Xi}) \bm{\alpha} + \bm{n},
    \label{eq:vector_model}
\end{equation}
where $\bm{\alpha} = [\alpha_1, \dots, \alpha_L]^T$ denotes a vector of all the multipath amplitudes, $\mathbf{A}(\bm{\Xi}) = [\bm{a}(\tau_1, \phi_1, \theta_1), \dots, \bm{a}(\tau_L, \phi_L, \theta_L)]$ is the matrix of the newly defined steering vectors for the parameter set $\bm{\Xi} = \{\bm{\xi}_1, \dots, \bm{\xi}_L\}$, and $\bm{n}$ is the noise vector. The model's compatibility with nonuniform sampling is explicitly embedded in the definition of the frequency set $\mathcal{F}$, which does not impose any structural constraints.

\subsection{MPC Extraction via the SAGE Algorithm} 
\label{subsec:sage}

The objective of MPC extraction is to estimate the complete set of path parameters $\bm{\Xi} = \{\bm{\xi}_1, \dots, \bm{\xi}_L\}$ from the measurement vector $\bm{y}$. Given the AWGN assumption in \eqref{eq:scalar_model}, the log-likelihood function of the observation vector $\bm{y}$ conditioned on the true parameters $\bm{\Xi}$ is derived as:
\begin{equation}
    \mathcal{L}(\bm{\Xi}) = -KM \log(\pi\sigma^2) - \frac{1}{\sigma^2} \left\| \bm{y} - \sum_{l=1}^{L} \alpha_l \bm{a}_l \right\|^2,
    \label{eq:log_likelihood}
\end{equation}
where for brevity we use $\bm{a}_l = \bm{a}(\tau_l, \phi_l, \theta_l)$, noting that this now represents the modified steering vector. The principle of maximum likelihood estimation (MLE) seeks the parameter set $\hat{\bm{\Xi}}$ that maximizes this function. A direct maximization of $L(\bm{\Xi} | \bm{y})$ is computationally intractable due to the high dimensionality and the non-linear dependence on the delay and DoA parameters. To circumvent this, we employ the SAGE algorithm, which decouples the multi-parameter problem into a sequence of simpler, single-path estimation subproblems. This problem decomposition facilitates the use of the nonuniform frequency sampling scheme as it avoids inference among the MPCs compared with direct IDFT, which will be shown in Sec.~VII-B. At the $i$-th iteration, to update the parameters of the $l$-th path, SAGE performs\cite{SAGE,DSS-SAGE}:
\begin{enumerate}
    \item \textbf{E-Step (Expectation)}: An interference-canceled observation, or ``target signal'', $\bm{y}_l^{(i)}$ for the $l$-th path is computed by subtracting the estimated contributions of all other paths from the total observation:
    \begin{equation}
        \bm{y}_l^{(i)} = \bm{y} - \sum_{j \neq l} \hat{\alpha}_j^{(i-1)} \bm{a}(\hat{\tau}_j^{(i-1)}, \hat{\phi}_j^{(i-1)}, \hat{\theta}_j^{(i-1)})
        \label{eq:sage_e_step}.
    \end{equation}
    \item \textbf{M-Step (Maximization)}: The parameter set $\bm{\xi}_l$ is updated by minimizing the corresponding likelihood function:
    \begin{equation}
        \hat{\bm{\xi}}_l^{(i)} = \arg \min_{\bm{\xi}_l} \mathcal{L}_l\triangleq \left\| \bm{y}_l^{(i)} - \alpha_l \bm{a}(\tau_l, \phi_l, \theta_l) \right\|^2.
    \end{equation}
\end{enumerate}
\par For a given set of delay and DoA parameters $(\tau, \phi, \theta)$, the complex amplitude $\alpha$ that minimizes the cost is found by projecting the target signal onto the steering vector $\bm{a} = \bm{a}(\tau, \phi, \theta)$. The optimal amplitude estimate is:
\begin{equation}
    \hat{\alpha}(\tau, \phi, \theta) = \frac{\bm{a}^H \bm{y}_l^{(i)}}{\|\bm{a}\|^2}.
\end{equation}
Substituting this optimal $\hat{\alpha}$ back into the likelihood function, we find that minimizing the cost is equivalent to maximizing the following objective function with respect to the non-linear parameters:
\begin{equation}
    \{\hat{\tau}_l, \hat{\phi}_l, \hat{\theta}_l\}^{(i)} = \arg \max_{\tau, \phi, \theta} \mathcal{L}_l\triangleq \left( \frac{\left| \bm{a}(\tau, \phi, \theta)^H \bm{y}_l^{(i)} \right|}{\|\bm{a}(\tau, \phi, \theta)\|}\right)^2.
    \label{eq:periodogram}
\end{equation}
The denominator is the squared norm of the steering vector:
\begin{align}
    &\|\bm{a}(\tau, \phi, \theta)\|^2 \\
    &= \left( \sum_{m=1}^{M} |G_{\text{ant}}(\phi, \theta | \bm{\Omega}_m)|^2 \right) \left( \sum_{k=1}^{K} |G_{\text{MA}}(f_k, \tau)|^2 \right).
\end{align}
Unlike models ignoring molecular absorption, the term $\sum_{k=1}^{K} |G_{\text{MA}}(f_k, \tau)|^2$ now introduces a dependency of the denominator on the delay parameter $\tau$. This dependency has a significant implication: the simplification of ignoring the denominator during the search for the optimal delay is no longer valid. The maximization in \eqref{eq:periodogram} must be performed over the complete, normalized expression via an explicit grid search over $\tau$.

To analyze the fundamental performance limits of delay estimation, we define the \textbf{single-path likelihood profile} as the normalized objective function, evaluated at the true DoA and viewed as a function of a test delay $\tau$, given by:
\begin{equation}
    \mathcal{L}_{\text{single}}(\tau) \triangleq \frac{ \bm{a}(\tau,  \hat{\phi}_l, \hat{\theta}_l)^H \cdot\bm{y}_{\text{target}} }{\|\bm{a}(\tau, \hat{\phi}_l, \hat{\theta}_l)\|}.
    \label{eq:likelihood_profile}
\end{equation}
where $\bm{y}_{\text{target}}=\bm{a}(\tau_\text{true},\phi_\text{true},\theta_\text{true})$ represent the ideal interference-canceled signal for a path with normalized amplitude obtained from the E-step, and $\tau_\text{true},\phi_\text{true},\theta_\text{true}$ denote the ground-truth delay and DoA of this path. $(\hat{\phi}_l, \hat{\theta}_l)$ is the estimated DoA of the path of interest. The properties of this function, including its mainlobe width, sidelobe levels, and ambiguous structure, determine the performance of the delay estimation. The primary focus of this paper is to analyze how different frequency sampling strategies and molecular absorption effect shape this fundamental likelihood profile. 

\begin{figure*}[ht]
    \centering 
        \centering
    \subfloat[UFS without MA]{%
        \includegraphics[width=0.3\textwidth]{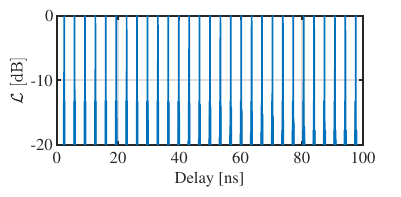}%
    }    
    \subfloat[CFS without MA]{%
        \includegraphics[width=0.3\textwidth]{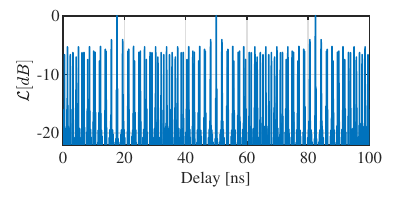}%
    }    
    \subfloat[NFS without MA]{%
        \includegraphics[width=0.3\textwidth]{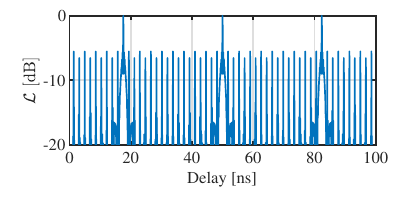}%
    }    
      \vspace{-0.2cm} 
    \subfloat[UFS with MA]{%
        \includegraphics[width=0.3\textwidth]{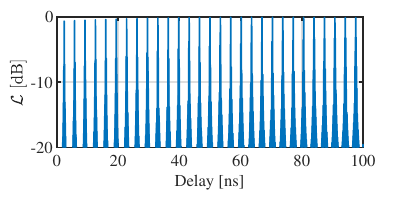}%
    }    
    \subfloat[CFS with MA]{%
        \includegraphics[width=0.3\textwidth]{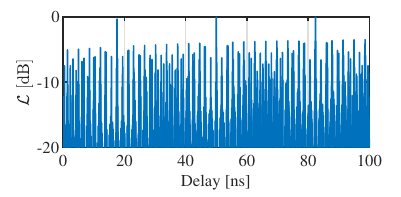}%
    }    
    \subfloat[NFS with MA]{%
        \includegraphics[width=0.3\textwidth]{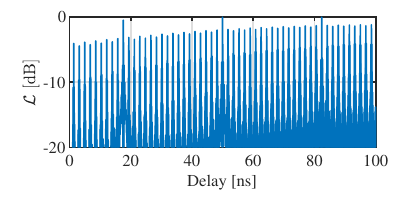}%
    }
    \caption{$\mathcal{L}_{\text{single}}(\tau)$ for different frequency sampling schemes and molecular absorption (MA) conditions with $\tau_\text{true}=50$~ns ($f_\text{c}=380~\text{GHz}$, $B=10~\text{GHz}$, and $K=35$).}
    \label{fig:three_subs}
    \vspace{-0.2cm} 
\end{figure*}

\section{Analysis and Comparison of Uniform, Coprime and Nested Frequency Sampling Schemes}
\label{sec:likelihood_analysis}

In this section, we analyze the topographical properties of the single-path likelihood profile defined in \eqref{eq:likelihood_profile} for the most common sampling schemes, i.e., uniform frequency sampling, coprime frequency sampling (CFS) and nested Frequency sampling (NFS). For the analysis of the likelihood profile with respect to a delay offset $\Delta\tau = \tau - \tau_\text{true}$, we first consider the ideal likelihood profile $\mathcal{L}_{\text{ideal}}(\Delta\tau)$ which does not consider molecular absorption effect. The effect of $G_{\text{MA}}(f, \tau)$ can be understood as a smoothing or broadening of the fundamental characteristics (resolution, ambiguity) of the ideal likelihood profile, $\mathcal{L}_{\text{ideal}}(\Delta\tau)$.

\subsection{Uniform Frequency Sampling}
\label{subsec:ufs}
We begin with the conventional uniform frequency sampling (UFS) scheme, which serves as a benchmark. The set of $K$ frequency points $\mathcal{F}$ is characterized by a constant frequency step $\Delta f$ over a total bandwidth $B = (K-1)\Delta f$. For a single-path, noise-free scenario, the ideal likelihood profile $\mathcal{L}_{\text{ideal, UFS}}(\Delta\tau)$ is dominated by the frequency-dependent sum of phasors yielding the well-known Dirichlet kernel as shown in Fig.~\ref{fig:three_subs}(a):
\begin{align}
    \mathcal{L}^{\text{UFS}}_{\text{ideal}}(\Delta\tau) \propto  \frac{\sin(\pi K \Delta f \Delta\tau)}{\sin(\pi \Delta f \Delta\tau)} .
    \label{eq:dirichlet_kernel}
\end{align}

The actual likelihood profile $\mathcal{L}_{\text{UFS}}(\Delta\tau)$ is a smoothed version of this function due to the convolution with $G_{\text{MA}}(\Delta\tau)$.

\subsubsection{Mainlobe Width and Resolution}
The delay resolution is dictated by the mainlobe width. The first nulls of the Dirichlet kernel occur at $|\Delta\tau| = 1/(K\Delta f) = 1/B$. The delay resolution, $\delta\tau$, is typically defined as half of the null-to-null width as $\delta\tau_{\text{UFS}} = 1/B$. The convolution with $G_{\text{MA}}(\Delta\tau)$ will cause a slight broadening of this mainlobe.

\subsubsection{Unambiguous Delay Range}
The periodic nature of the Dirichlet kernel gives rise to ambiguous peaks when the denominator in \eqref{eq:dirichlet_kernel} approaches zero, which occurs at delay offsets of $\Delta\tau_{\text{amb}} = n/\Delta f$ as shown in Fig.~\ref{fig:three_subs}(a). The unambiguous delay range (UDR) is the first peak as $T_{\text{amb, UFS}} = \frac{1}{\Delta f}$ under the constraint of a fixed bandwidth $B$ and $K$ samples. The rigid relationship $\Delta f = B/(K-1)$ allows us to express the UDR as:
\begin{equation}
    \text{UDR}_{\text{UFS}} = \frac{1}{\Delta f} = \frac{K-1}{B}.
    \label{eq:udr_ufs_final}
\end{equation}

For a large number of points ($K \gg 1$), this can be approximated as $\text{UDR}_{\text{UFS}} \approx K/B$. This expression reveals a direct, linear dependence of the UDR on the number of samples, highlighting the inflexible trade-off of the UFS method.

\subsection{Coprime Frequency Sampling}
\label{subsec:cfs}
Coprime frequency sampling (CFS) utilizes two sparse uniform sub-arrays to synthesize a large virtual bandwidth. Let $M$ and $N$ be a pair of coprime integers (where $M < N$ and $\text{gcd}(M,N)=1$). The CFS frequency set $\mathcal{F}_{\text{coprime}}$ is constructed as the union of two sub-sets, $\mathcal{S}_M$ and $\mathcal{S}_N$, sharing a fundamental frequency step $\Delta f_{\text{base}}$:
\begin{subequations}
\begin{align}
    \mathcal{S}_M &= \{ m \cdot (N \Delta f_{\text{base}}) \mid 0 \le m \le M-1 \}, \\
    \mathcal{S}_N &= \{ n \cdot (M \Delta f_{\text{base}}) \mid 0 \le n \le N-1 \}, \\
    \mathcal{F}_{\text{coprime}} &= \mathcal{S}_M \cup \mathcal{S}_N.
\end{align}
\end{subequations}
The total number of samples is $K = M + N$. The ideal likelihood profile is the superposition of the responses from these two sparse sub-arrays. Consequently, \eqref{eq:cfs_sum} can be explicitly rewritten to reflect this dual-structure:
\begin{equation}
    \mathcal{L}^{\text{CFS}}_{\text{ideal}}(\Delta\tau) \propto \underbrace{\sum_{m=0}^{M-1} e^{j 2\pi (m N \Delta f_{\text{base}}) \Delta\tau}}_{\text{Sub-array } M} + \underbrace{\sum_{n=1}^{N-1} e^{j 2\pi (n M \Delta f_{\text{base}}) \Delta\tau}}_{\text{Sub-array } N}.
    \label{eq:cfs_sum}
\end{equation}
The first term corresponds to a sub-array with coarse spacing $N\Delta f_{\text{base}}$, while the second has spacing $M\Delta f_{\text{base}}$.

\subsubsection{Unambiguous Delay Range}
The UDR in CFS arises from the non-alignment of the grating lobes between the two sub-arrays. Due to the coprime property of $M$ and $N$, the joint ambiguity occurs only when the peaks of both sub-arrays align. This alignment happens at the least common multiple (LCM) of their periods:
\begin{equation}
    \text{UDR}_{\text{CFS}} = \text{LCM}\left(\frac{1}{N \Delta f_{\text{base}}}, \frac{1}{M \Delta f_{\text{base}}}\right) = \frac{1}{\Delta f_{\text{base}}}.
\end{equation}
This implies that the effective UDR is determined by the fundamental step $\Delta f_{\text{base}}$, despite the sparse sampling.
To quantify the gain, we express the total bandwidth $B$ spanned by the coprime array. Since the two sub-arrays cover a spectral range up to approximately $MN$ times the base step, we have:
\begin{equation}
     \Delta f_{\text{base}} = \frac{B}{MN}.
\end{equation}
Substituting $\Delta f_{\text{base}}$ into the UDR expression yields $\text{UDR}_{\text{CFS}} = \frac{MN}{B}$.
Given a fixed resource of samples $K \approx M+N$, the goal is to maximize the product $MN$ subject to the coprime constraint $\gcd(M,N)=1$.
For an odd integer $K$, the optimal allocation is achieved by choosing two consecutive integers, $M = (K-1)/2$ and $N = (K+1)/2$. Since the two consecutive integers are inherently coprime, this configuration ensures validity while maximizing the product to $MN = (K^2-1)/4$. Consequently, the theoretical maximum UDR is:
\begin{equation}
    \text{UDR}_{\text{CFS}} = \frac{K^2 - 1}{4B} \approx \frac{K^2}{4B}.
    \label{eq:udr_cfs_final}
\end{equation}
Comparison with uniform sampling implies that CFS expands the unambiguous delay range by a factor of approximately $K/4$.

\subsubsection{Sidelobe Properties}
The CFS profile exhibits a low sidelobe floor, but is punctuated by deterministic peaks arising from the periodicity of the constituent sub-arrays. These peaks occur at offsets $\tau_{\text{sidelobe}, M} = n/(N\Delta f_{\text{base}})$ and $\tau_{\text{sidelobe}, N} = n/(M\Delta f_{\text{base}})$. As a result, the sidelobes are observed nonuniform in the delay domain in the likelihood function as shown in Fig.~\ref{fig:three_subs}(b). Sidelobe floor is approximately -6 dB relative to the main peak, often defining the practical dynamic range of the scheme. This occurs because the whole array is divided into two sub-arrays, resulting in that the amplitude of those sidelobes is approximately half of the mainlobe.

\subsection{Nested Frequency Sampling}
\label{subsec:nfs}
Nested frequency sampling (NFS) employs a hierarchical non-uniform strategy to extend the aperture by concatenating a dense inner array with a sparse outer array. An NFS set $\mathcal{F}_{\text{nested}}$ is formed by the union of two distinct uniform sub-sets, $\mathcal{S}_{\text{dense}}$ and $\mathcal{S}_{\text{sparse}}$, sharing a fundamental frequency step $\Delta f_{\text{base}}$.
Let $N_1$ and $N_2$ denote the number of samples in the dense and sparse sub-arrays, respectively. The frequency points are generated as:
\begin{subequations}
\begin{align}
    \mathcal{S}_{\text{dense}} &= \{ n \cdot \Delta f_{\text{base}} \mid 1 \le n \le N_1 \}, \\
    \mathcal{S}_{\text{sparse}} &= \{ m \cdot N_1  \Delta f_{\text{base}} \mid 1 \le m \le N_2 \}, \\
    \mathcal{F}_{\text{nested}} &= \mathcal{S}_{\text{dense}} \cup \mathcal{S}_{\text{sparse}}.
\end{align}
\end{subequations}
The total number of sampling points is $K = N_1 + N_2$. The ideal likelihood profile is the superposition of the responses from these two sub-arrays, which can be expressed as:
\begin{equation}
\begin{split}
        &\mathcal{L}^{\text{UFS}}_{\text{ideal}}(\Delta\tau)
        \\&\propto \underbrace{\sum_{n=1}^{N_1} e^{j 2\pi (n \Delta f_{\text{base}}) \Delta\tau}}_{\text{Dense Sub-array}} + \underbrace{\sum_{m=1}^{N_2} e^{j 2\pi (mN_1 \Delta f_{\text{base}}) \Delta\tau}}_{\text{Sparse Sub-array}}.
\end{split}
\end{equation}
The first term corresponds to the inner array with fine spacing $\Delta f_{\text{base}}$, while the second term represents the outer extension with coarse spacing $N_1\Delta f_{\text{base}}$.

\subsubsection{Unambiguous Delay Range}
The UDR of the NFS scheme is governed by the dense sub-array. Since the positions in $\mathcal{S}_{\text{dense}}$ are consecutive multiples of $\Delta f_{\text{base}}$, the greatest common divisor of the entire set is exactly $\Delta f_{\text{base}}$. Consequently, the rigorous ambiguity period is $\text{UDR}_{\text{NFS}} = \frac{1}{\Delta f_{\text{base}}}$. The total bandwidth $B$ covered by the nested array extends up to the last element of the sparse set, the UDR expression yields:
\begin{equation}
    \text{UDR}_{\text{NFS}} = \frac{N_2N_1}{B}.
\end{equation}
According to the arithmetic-geometric mean inequality, the product is maximized when $N_1 = N_2 = K/2$.
Substituting these optimal values back into the equation results in the theoretical maximum UDR:
\begin{equation}
    \text{UDR}_{\text{NFS}} = \frac{(K/2)(K/2)}{B} = \frac{K^2}{4B}.
    \label{eq:udr_nfs_final}
\end{equation}
This confirms that, similar to CFS, the NFS scheme achieves an $O(K^2)$ expansion in the UDR compared to the $O(K)$ scaling of uniform sampling.

\subsubsection{Sidelobe Properties}
The hole-free difference co-array of the NFS scheme results in a flatter statistical sidelobe floor compared to CFS. However, a deterministic sidelobe peak still exists, which is generated by the periodicity of the sparse sub-array at a delay offset of $\tau_{\text{sidelobe, NFS}} = 1/(N_1\Delta f_{\text{base}})$. For an optimal design, the power of this peak is also approximately -6 dB relative to the mainlobe and often defines the effective dynamic range, which is due to the same reason as CFS.
\subsection{Performance Comparison of Unambiguous Delay Range}
\label{subsec:udr_comparison}
The derivations above provide explicit expressions for the maximum achievable UDR for each scheme under the constraints of a fixed bandwidth $B$ and a total of $K$ sampling points. All three expressions share a common factor of $K/B$, which can be interpreted as the baseline UDR provided by a UFS scheme. The nonuniform sampling schemes, CFS and NFS, enhance this baseline by a multiplicative ``gain factor'' calculated as $K/4$. Taking the bandwidth of 60~GHz and maximum delay of 200~ns as an example, 12000 frequency samples are needed for UFS, and by comparison, this number is 220 for nonuniform frequency sampling schemes, CFS and NFS. This implies that in the THz channel sounding with large bandwidth, a frequency sampling scheme with a large UDR will reduce the channel sounding time. However, both coprime and nested schemes, while extending the unambiguous range, still exhibit unavoidable ambiguous lobes (sidelobes at -6 dB level). Hence, a new scheme that completely eliminates delay ambiguity is needed.
\subsection{Distortion on $\mathcal{L}_{\text{idea}}(\tau)$}
Figure~\ref{fig:three_subs} shows the likelihood functions of a 50~ns path delay for UFS, CFS and NFS with and without molecular absorption effects. The central frequency is set as 380~GHz with the bandwidth of 10~GHz. The temperature is set to be $30^\circ \mathrm{C}$ and water vapor density is 20~g/m3. The comparisons in the figure reveals that the molecular absorption could distort the likelihood function, including broadening the mainlobe and enhancing the sidelobes. A broadened mainlobe would lead to a reduced precision in estimating the delay. Also, enhanced sidelobes may disturb delay estimation once the inference from other MPCs is not ideally canceled in the E step. This implies that an undistorted likelihood function is necessary for the delay estimation based on SAGE algorithm.
\section{Delay-Ambiguity-Free Sampling Scheme for Channel Sounding}
This section develops a novel frequency sampling scheme that eliminates delay ambiguity, a key limitation of existing wideband sounding methods. We first formulate a mathematical framework based on the Poisson Summation Formula to explicitly decompose the channel's delay likelihood function into a mainlobe and a series of ambiguous lobes. By analyzing this decomposition with the Method of Stationary Phase, we derive fundamental design principles for ambiguity-free sampling. Finally, we propose and detail a Parabolic Frequency Sampling (PFS) scheme that adheres to these principles.
\subsection{Explicit Analysis of Single-path Likelihood Function}
The likelihood function, as defined in \eqref{eq:likelihood_profile}, is proportional to the squared magnitude of a function that we will denote as $\mathcal{L}_\text{single}(\tau)$. We now proceed to analyze this function using the Poisson Summation Formula to understand how the path gain $G_{\text{MA}}(f, \tau)$ and the frequency sampling scheme jointly shape the profile.

Let us consider the single-path likelihood function with respective to $\bm{y}_{\text{target}}$ in \eqref{eq:likelihood_profile}. To analyze its structure, we consider an ideal, noise-free, single-path scenario where the target signal $\bm{y}_{\text{target}}$ corresponds to a single true path with parameters $\bm{\xi}_{\text{true}} = \{\alpha_{\text{true}}, \tau_{\text{true}}, \hat{\phi}_l, \hat{\theta}_l\}$. In this case, the target signal is $\bm{y}_{\text{target}} = \alpha_{\text{true}} \bm{a}(\tau_{\text{true}}, \hat{\phi}_l, \hat{\theta}_l)$. Substituting this into the definition of $\mathcal{L}_\text{single}(\tau)$, we obtain:
\begin{align}
    \mathcal{L}_\text{single}(\tau)= \frac{\alpha_{\text{true}} \sum_{m=1}^{M} \sum_{k=1}^{K} [\bm{a}(\tau, \dots)]_{k,m}^* [\bm{a}(\tau_{\text{true}}, \dots)]_{k,m}}{{\|\bm{a}(\tau, \hat{\phi}_l, \hat{\theta}_l)\|}},
\end{align}
where the asterisk denotes the complex conjugate. Inserting the definition of the steering vector from Section \ref{sec:system_model}, and ignoring the spatial term, $\sum_{m=1}^{M} |G_{\text{ant}}(\hat{\phi}_l, \hat{\theta}_l | \bm{\Omega}_m)|^2$, which is a constant scaling factor, the expression for $\mathcal{L}_\text{single}(\tau)$ simplifies to:
\begin{equation}
    \mathcal{L}_\text{single}(\tau) \propto \sum_{k=1}^{K} \frac{G_{\text{MA}}(f_k, \tau)G_{\text{MA}}(f_k, \tau_\text{true}) }{{\|\bm{a}(\tau, \hat{\phi}_l, \hat{\theta}_l)\|}}e^{j 2\pi f_k (\tau-\tau_\text{true}))}.
\end{equation}

To apply the Poisson Summation Formula, we generalize this discrete sum to a continuous framework. We represent the discrete frequencies $f_k$ as samples of a continuous function $f(v)$, where $v$ is a continuous frequency index. The summation can then be viewed as samples of a continuous function $h(v;\tau)$ at integer points $v=k$:
\begin{equation}
    h(v; \tau) \triangleq \frac{ G_{\text{MA}}(f(v), \tau)G_{\text{MA}}(f(v), \tau_\text{true}) }{{\|\bm{a}(\tau, \hat{\phi}_l, \hat{\theta}_l)\|}}e^{j 2\pi f(v)\tau}.
\end{equation}

The discrete sum $\mathcal{L}_\text{single}(\tau)$ is thus represented as $\sum_{k=1}^{K} h(k; \Delta\tau)$. Applying the Poisson Summation Formula~\cite{PSF}, $\sum_{n=-\infty}^{\infty} h(n) = \sum_{m=-\infty}^{\infty} H(2\pi m)$, where $H(\omega)$ is the continuous-time Fourier transform of $h(v)$ with respect to $v$, we can express $\mathcal{L}_\text{single}(\tau)$ as a sum over its spectral components:

\begin{equation}
    \mathcal{L}_\text{single}(\tau) = \sum_{m=-\infty}^{\infty} s_m(\tau),
\end{equation}
where each component $s_m(\Delta\tau)$ is the $m$-th term of the Fourier series representation, given by:
\begin{align}
    s_m(\tau) =& H(2\pi m; \tau) = \int_{-\infty}^{\infty} h(v; \tau) e^{-j 2\pi m v} dv \nonumber \\
    =& \int_{-\infty}^{\infty} \frac{G_{\text{MA}}(f(v), \tau)G_{\text{MA}}(f(v), \tau_\text{true}) }{{{\|\bm{a}(\tau, \hat{\phi}_l, \hat{\theta}_l)\|}}}    \label{eq:sm_tau_final}\\
    &\quad\cdot e^{j 2\pi f(v) (\tau - \tau_{\text{true}})}e^{-j 2\pi m v} dv.\nonumber
\end{align}

The equation~\eqref{eq:sm_tau_final} is the fundamental result of this analysis. It decomposes the complex response $\mathcal{L}_\text{single}(\tau)$ into a series of components. The term, $s_0(\tau)$ ($m=0$), represents the mainlobe of the likelihood profile, shaped by both the sampling scheme $f(v)$ and the frequency-dependent path gain $G_{\text{MA}}$. The terms $s_m(\tau)$ for $m \neq 0$ represent the ambiguous lobes (or platforms), which are also modulated by the path gain. This formulation allows for a powerful analysis of how $f(v)$ and $G_{\text{MA}}$ impacts the resolution (via $s_0$) and ambiguous structure (via $s_{m\neq0}$) for any given frequency sampling strategy $f(v)$.
\subsection{Design Principles}

The decomposition in \eqref{eq:sm_tau_final} is fundamental to understanding and controlling delay ambiguities. The terms $s_m(\tau)$ for $m \neq 0$ represent the ambiguous lobes, and our goal is to design a system where these terms are minimized. For a large bandwidth, the exponential term in the integral for $s_m(\tau)$ is a rapidly oscillating function of the integration variable $v$. For such integrals, the Method of Stationary Phase (MSP) provides a powerful approximation, stating that the main contribution to the integral comes from points where the phase changes most slowly~\cite{PSF,SFCW-PSF}.

Let us define the phase term of the integrand in \eqref{eq:sm_tau_final} as $\Phi_m(v; \tau)$:
\begin{equation}
    \Phi_m(v; \tau) \triangleq 2\pi \left( f(v)(\tau - \tau_{\text{true}}) - mv \right).
\end{equation}
The stationary phase condition requires that the derivative of the phase with respect to $v$ is zero. Let $f'(v) = df(v)/dv$ denote the rate of change of frequency with respect to the index, i.e., the local frequency step. Setting the derivative of $\Phi_m(v; \tau) $ to zero, we find that the delay of lobes satisfies:
\begin{equation}
    \tau=\tau_{\text{true}}+\frac{m}{f'(\tilde{v})}.
        \label{eq:ambiguous_location}
\end{equation}
This relationship is the key to designing an ambiguity-free scheme and gives rise to two fundamental design principles.
\par First, it shows that if $f'(v)$ is not a constant but varies with $v$, it will form ambiguous peaks at different delay offsets. This will spread the power of sidelobes to a rand range rather than focus on the sidelobes on the fixed and periodic delays as UFS which results in delay ambiguity shown in Fig.~\ref{fig:three_subs}(a). Second, to avoid overlapping of two adjacent sidelobes which would cause an enhanced sidelobe power, the following condition must be satisfied:
\begin{equation}
    \max\{\frac{m}{f'(v)}\}\geq\min\{\frac{m+1}{f'(v)}\},
\end{equation}
which results in a constraint of $\min\{f'(v)\}\geq \max\{f'(v)\}/2$ when $m=1$.
Therefore, a delay-ambiguity-free frequency sampling scheme can be realized by designing the frequency sampling function $f(v)$. The design principles are twofold:
\begin{enumerate}
        \item \textbf{A nonuniform frequency sampling:} design $f(v)$ to be a non-linear function, ensuring the local frequency step $f'(v)$ varies significantly over the integration interval, which spreads the sidelobes energy.
        \item \textbf{A proper minimum value of $f'(v)$:} make sure that the minimum value of the local frequency step $f'(v)$ is larger than a half of its maximum value.
\end{enumerate}

\subsection{Proposed Parabolic Frequency Sampling Scheme}
\label{subsec:pds}
Guided by the principles established above, we now propose a practical delay-ambiguity-free scheme. The first principle requires a non-linear $f(v)$. A quadratic function for the local frequency step $f'(v)$ is the simplest non-trivial form that satisfies this requirement. This choice is also practically advantageous, as it concentrates frequency samples in the center of the band, which can improve the signal-to-noise ratio in many scenarios. We term this the Parabolic Frequency Sampling (PFS) scheme.

We define the local frequency step $f'(v)$ for a set of $K$ samples (indexed $v \in [1, K]$) as a parabola symmetric about the midpoint $(K+1)/2$:
\begin{equation}
    f'(v) = a \left(v - \frac{K+1}{2}\right)^2 + \kappa,
    \label{eq:pds_derivative}
\end{equation}
where the curvature $a > 0$ and the minimum step $\kappa \ge 0$. The parameters $a$ and $\kappa$ are uniquely determined by the system specifications (total bandwidth $B$ and number of samples $K$) and our second design principle. The maximum and minimum frequency steps are:
\begin{align}
    f'_{\text{max}} &= f'(1) = a \left(\frac{K-1}{2}\right)^2 + \kappa, \\
    f'_{\text{min}} &= f'\left(\frac{K+1}{2}\right) = \kappa.
\end{align}
We now enforce the second design principle by setting $f'_{\text{min}} = \frac{1}{2} f'_{\text{max}}$. This yields a direct relationship between the parameters:
\begin{equation}
    \kappa = a \frac{(K-1)^2}{4}.
    \label{eq:kappa_a_relation}
\end{equation}
The parameters must also satisfy the total bandwidth constraint, which requires that the integral of the frequency step over the index range equals $B$:
\begin{equation}
    \int_{1}^{K} f'(v) \,dv = \int_{1}^{K} \left[ a \left(v - \frac{K+1}{2}\right)^2 + \kappa \right] dv = B.
    \label{eq:pds_integral_constraint}
\end{equation}
We have $a \frac{(K-1)^3}{12} + \kappa(K-1) = B$. Solving the system of equations formed by \eqref{eq:kappa_a_relation} and this bandwidth constraint uniquely determines $a$ and $\kappa$:
\begin{equation}
\begin{split}
    a = \frac{3B}{(K-1)^3},
    \\ \kappa = \frac{3B}{4(K-1)}.
    \end{split}
\end{equation}
\par
\begin{figure}[t]
\centering
    \subfloat[$f'(k)$]{%
        \includegraphics[width=0.4\textwidth]{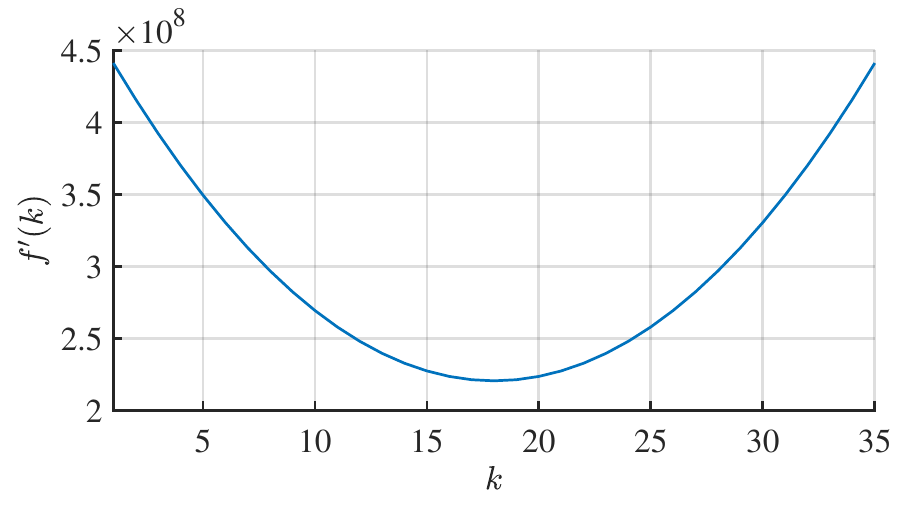}%
    }
    
    \subfloat[$f(k)$]{%
        \includegraphics[width=0.4\textwidth]{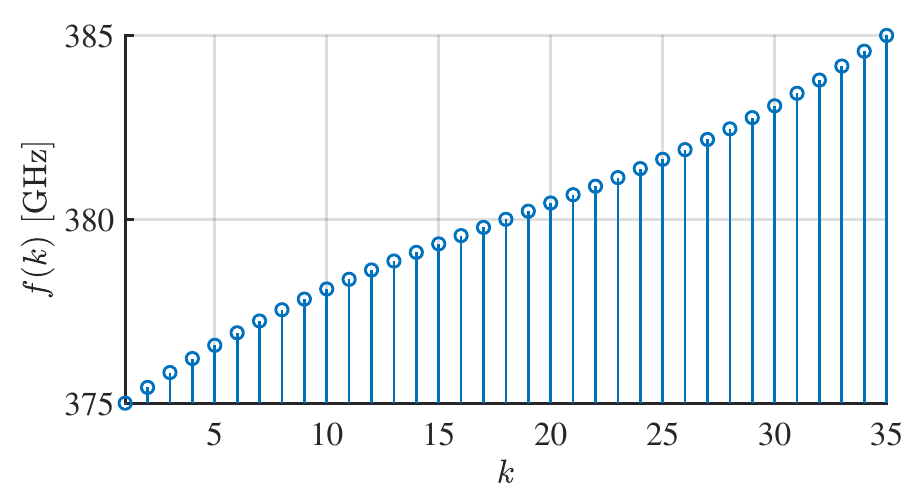}%
    }
    \caption{Proposed parabolic frequency sampling scheme. (a) Local frequency step, $f'(k)$, and (b) Frequency distribution function, $f(k)$. ($f_\text{c}=380~\text{GHz}$, $B=10~\text{GHz}$ and $K=35$).}
    \label{fig:PFS}
\end{figure}

 Once $a$ and $\kappa$ determined, the frequency distribution function $f(v)$ is obtained by integrating \eqref{eq:pds_derivative} and setting the initial condition $f(1) = f_{\text{start}}$:
\begin{align}
f_\text{PFS}(v)
&=  \frac{B}{(K-1)^3} \left[ \left(v - \frac{K+1}{2}\right)^3+\frac{(K-1)^3}{8} \right] \\
&+ \frac{3B(v-1)}{4(K-1)}+f_{\text{start}} .
\end{align}
This yields the final set of frequency points for the proposed parabolic frequency sampling scheme. Figure~\ref{fig:PFS} shows an example of the proposed parabolic frequency sampling scheme with central frequency of 380~GHz, bandwidth of 10~GHz, and a total frequency sampling number of 35. The sampled frequencies are densely located at the central frequency and well cover the whole interested spectrum.
\section{LR-SAGE Algorithm}
In this section, we introduce a LR-SAGE algorithm to to rectify the distortion in the likelihood function’s mainlobe caused by molecular absorption. First, we present the rectification principle for the likelihood function by introducing a post-processing factor. We derive a frequency- and delay-dependent factor $I(v; \tau)$ to compensate for the molecular attenuation $G_{MA}(f(v),\tau)$ and local frequency step $f'(v)$. This factor effectively sharpens the likelihood mainlobe. Second, we integrate this rectification factor into the SAGE algorithm’s M-step, yielding the LR-SAGE algorithm.
\subsection{Likelihood Mainlobe Rectification}
\par The frequency-dependent gain due to molecular absorption effect would distort the mainlobe and make the delay estimation inaccurate. A critical goal is to ensure that the shape of the mainlobe response, represented by $s_0(\tau)$, remains undistorted and equivalent to that of a dense Uniform Frequency Sampling system. UFS scheme produces an ideal likelihood shape in the delay domain, whose mainlobe width, and thus the system's resolution, is determined solely by $B$. We must ensure that our nonuniform scheme preserves this fundamental property and approaches to the benchmark response from a UFS system, which is equivalent to the inverse Fourier transform of a rectangular window of width $B$:
\begin{equation}
    s_{0, \text{UFS}}(\tau) \propto \int_{-B/2}^{B/2}e^{j 2\pi u \Delta\tau} du.
    \label{eq:s0_ufs_benchmark}
\end{equation}
\par We start by introducing a frequency- and delay-dependent factor $I(v; \tau)$ to the likelihood function. The integral for the mainlobe component given by setting $m=0$ in \eqref{eq:sm_tau_final} is now calculated as:
\begin{equation}
    s_0(\tau) = \int_{-\infty}^{\infty} \frac{I(v; \tau)G_{\text{MA}}(f(v), \tau) }{{\|\bm{a}(\tau, \hat{\phi}_l, \hat{\theta}_l)\|}} e^{j 2\pi f(v) (\tau - \tau_{\text{true}})} dv.
\end{equation}

Let $u = f(v) - f_\text{c}$, where $f_\text{c}$ is the center frequency. This maps the integration to an integration over the frequency variable $u$ spanning the bandwidth $[-B/2, B/2]$. The differential element transforms as $dv = du / f'(v)$, where $f'(v) = df(v)/dv$. The integral for $s_0(\tau)$ then becomes
\begin{equation}
        s_0(\tau) \propto \int_{-B/2}^{B/2} \frac{I(f^{-1}(u))G_{\text{MA}}(f^{-1}(u), \tau)}{f'(f^{-1}(u)){\|\bm{a}(\tau, \hat{\phi}_l, \hat{\theta}_l)\|}} e^{j 2\pi u \Delta\tau} du,
    \label{eq:s0_simplified}
\end{equation}
where $\Delta\tau = \tau - \tau_{\text{true}}$ and the integration is over the total bandwidth $[-B/2, B/2]$. For this integral to be equivalent to the ideal Fourier transform in \eqref{eq:s0_ufs_benchmark}, the term multiplying the exponential must be a constant over the entire integration interval. This gives us our second fundamental design constraint:
\begin{equation}
    \frac{I(v; \tau)G_{\text{MA}}(f(v), \tau)G_{\text{MA}}(f(v), \tau_\text{true})}{f'(v){\|\bm{a}(\tau, \hat{\phi}_l, \hat{\theta}_l)\|}} = \text{const.}
    \label{eq:constraint_fidelity}
\end{equation}
\par This crucial result shows that to maintain mainlobe fidelity, the post-processing gain $I(v)$ must be designed to be directly proportional to the local frequency step size $f'(v)$. For simplicity and without loss of generality, we can normalize this constant to one, yielding the elegant constraint:
\begin{equation}
    I(v;\tau) = \frac{f'(v){\|\bm{a}(\tau, \hat{\phi}_l, \hat{\theta}_l)\|}}{G_{\text{MA}}(f(v), \tau)G_{\text{MA}}(f(v), \tau_\text{true})}.
    \label{eq:fidelity_constraint_final}
\end{equation}
\par The expression for $I(v;\tau)$ reveals that the rectification of the distorted mainlobe of likelihood function depends on both the derivative of frequency distribution function $f'(v)$ and the path gain $G_{\text{MA}}(f, \tau)$ due to molecular absorption. That is, any nonuniform frequency sampling scheme, where $f'(v)$ is not a constant, is suggested to carry out the likelihood rectification.
\begin{algorithm}[ht]
\caption{LR-SAGE Algorithm}
\label{alg:sage}
\begin{algorithmic}[1] 
\STATE \textbf{Input:}
\STATE \quad Measurement vector $\bm{y} \in \mathbb{C}^{KM \times 1}$
\STATE \quad Initial number of paths $\hat{L}$, Initial path parameter estimates $\hat{\bm{\Xi}}^{(0)}$, Maximum number of iterations $I_{\text{max}}$, and Convergence threshold $\epsilon$.

\STATE \textbf{Initialization:}
\STATE \quad Set iteration counter $i \leftarrow 1$
\WHILE{$i \le I_{\text{max}}$ and $|\mathcal{L}^{(i-1)} - \mathcal{L}^{(i-2)}| > \epsilon$}
    \FOR{$l = 1, \dots, \hat{L}$}
        \STATE \textbf{E-Step}: Compute interference-cancelled signal.
        \STATE Compute target signal for the $l$-th path:
        \begin{equation*}
            \bm{y}_l^{(i)} = \bm{y} - \sum_{j=1, j \neq l}^{\hat{L}} \hat{\alpha}_j^{(i-1)} \bm{a}(\hat{\tau}_j^{(i-1)}, \hat{\phi}_j^{(i-1)}, \hat{\theta}_j^{(i-1)})
        \end{equation*}
        
        \STATE  \textbf{M-Step}: Update parameters for the $l$-th path.
        \STATE Update delay and DoA by maximizing the rectified likelihood function:
        \begin{align*}
            &\{\hat{\tau}_l^{(i)}, \hat{\phi}_l^{(i)}, \hat{\theta}_l^{(i)}\} 
            \\&= \arg \max_{\tau, \phi, \theta} \frac{\left| \bm{I}(\tau;\hat\tau^{(i-1)}_l)\bm{a}(\tau, \phi, \theta)^H \bm{y}_l^{(i)} \right|^2}{\|\bm{a}(\tau, \phi, \theta)\|^2}
        \end{align*}
        
        \STATE Form the updated steering vector for the $l$-th path:
        \STATE $\hat{\bm{a}}_l^{(i)} \leftarrow \bm{a}(\hat{\tau}_l^{(i)}, \hat{\phi}_l^{(i)}, \hat{\theta}_l^{(i)})$ and update complex amplitude, $\hat{\alpha}_l^{(i)}$ using eq. (8)
        
        \STATE Update the parameter set for the current path:
         $\hat{\bm{\xi}}_l^{(i)} \leftarrow \{\hat{\alpha}_l^{(i)}, \hat{\tau}_l^{(i)}, \hat{\phi}_l^{(i)}, \hat{\theta}_l^{(i)}\}$
    \ENDFOR
    
    \STATE Update the full parameter set: $\hat{\bm{\Xi}}^{(i)} \leftarrow \{\hat{\bm{\xi}}_1^{(i)}, \dots, \hat{\bm{\xi}}_{\hat{L}}^{(i)}\}$
    \STATE $i \leftarrow i + 1$
\ENDWHILE
\STATE \textbf{Output:}
\STATE \quad Final estimated path parameters $\hat{\bm{\Xi}} = \hat{\bm{\Xi}}^{(i-1)}$
\end{algorithmic}
\end{algorithm}
\subsection{Procedure of the LR-SAGE Algorithm}
The main modification to the original SAGE algorithm is the M-step of the original SAGE algorithm. The optimization goal in \eqref{eq:periodogram} is modified by introducing an additional post-processing factor as:
\begin{equation}
        \{\hat{\tau}_l, \hat{\phi}_l, \hat{\theta}_l\}^{(i)} = \arg \max_{\tau, \phi, \theta} \frac{\left| (\bm{I}(\tau;\hat\tau^{(i-1)}_l)\cdot\bm{a}(\tau, \phi, \theta))^H \bm{y}_l^{(i)} \right|^2}{\|\bm{a}(\tau, \phi, \theta)\|^2},
    \label{eq:modified}
\end{equation}
where $\bm{I}(\tau;\hat\tau^{(i-1)}_l)$ is a vectorized $I(f;\tau)$ expressed in \eqref{eq:fidelity_constraint_final}. If we assume $\tau_\text{true}$ for the current path is the estimated path delay in the last round, $\bm{I}(\tau;\hat\tau^{(i-1)}_l)$ is denoted as:
\begin{equation}
    \bm{I}(\tau;\hat\tau^{(i-1)}_l)=[I(f_1,\tau), \dots, I(f_K,\tau)]^T |_{\tau_\text{true}=\hat\tau^{(i-1)}_l}.
\end{equation}

The complete LR-SAGE algorithm involving the above modification is summarized in Algorithm \ref{alg:sage}.
\par A critical step in this LR-SAGE algorithm is to calculate the molecular absorption path gain $G_{\textbf{MA}}(f_k,\tau)$ given the frequency and testing delay in \eqref{eq:fidelity_constraint_final}, which is in practice difficult. If the channel sounding is conducted in a spectrum where molecular absorption is not severe, one can simply set this path gain as one. Thus, the LR-SAGE algorithm focuses on dealing with the distortion caused by nonuniform frequency sampling. However, if a spectrum is highly affected by molecular absorption, we also present a simple method to estimate $G_{\textbf{MA}}(f_k,\tau)$ based on the Beer-Lambert law~\cite{multi-ray}:
\begin{equation}
    \log{G_{\textbf{MA}}(f_k,\tau)}=\frac{c\tau\log{G^m_{\textbf{MA}}(f_k,d_\text{ref}/c)}}{d_\text{ref}},
\end{equation}
where $d_\text{ref}$ is the reference distance for measuring, $G^m_{\textbf{MA}}(f_k,d_\text{ref}/c)$ is the path gain measured at frequency $f_k$ and the reference distance.

\section{Simulation Results}
In this section, we use simulation to verify the proposed parabolic frequency sampling scheme and the LR-SAGE algorithm. First, we examine the single-path likelihood function of the proposed PFS and the performance of rectification. Next, we use a synthesized multipath channel to compare the performance of the LR-SAGE algorithm under different frequency sampling schemes.
 \begin{figure}[!ht]
\centering
    \subfloat[$K=20$]{%
        \includegraphics[width=0.45\textwidth]{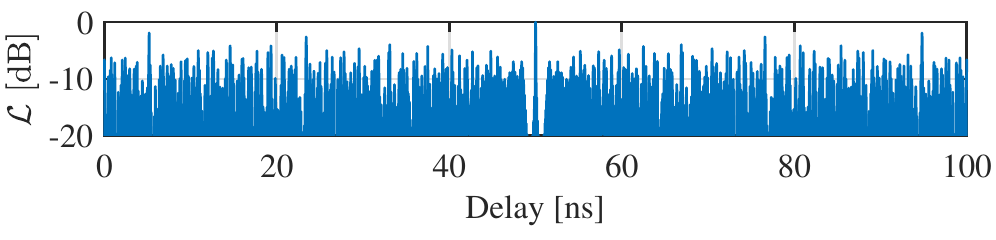}%
    }

    \subfloat[$K=35$]{%
        \includegraphics[width=0.45\textwidth]{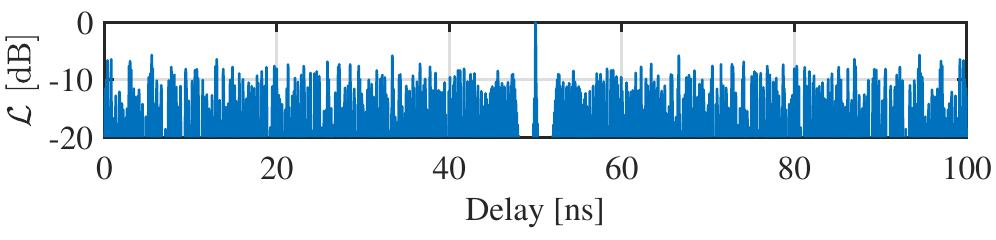}%
    }

    \subfloat[$K=50$]{%
        \includegraphics[width=0.45\textwidth]{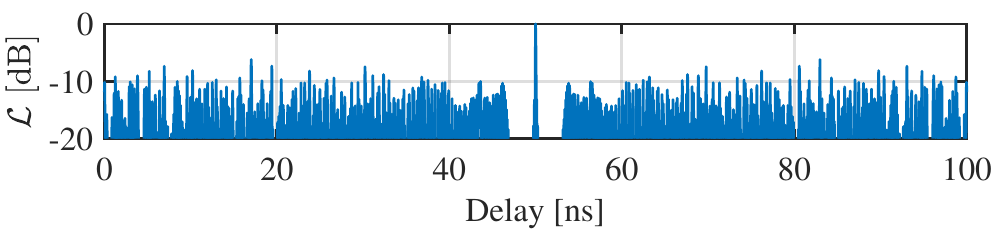}%
    }
    \caption{Rectified Likelihood function $\mathcal{L}(\tau)$ of the proposed parabolic frequency sampling scheme for a path delay of 50~ns with (a) $K=20$, (b) $K=35$ and (c) $K=50$  ($f_\text{c}=375~\text{GHz}$, $B=10~\text{GHz}$, $T=30^\circ \mathrm{C}$ and $D_\text{VP}=20~\text{g/m}^3$).}
    \label{fig:L-K}
\end{figure}
\subsection{Likelihood Function of Proposed Parabolic Sampling Scheme}
 In Fig. ~\ref{fig:L-K}, we calculate the rectified single-path likelihood function $\mathcal{L}(\tau)$ of the proposed parabolic sampling scheme with a path delay of 50~ns and  different total numbers of frequency samples $K$. The central frequency is set as 375 GHz with a bandwidth of 10~GHz. It shows that even when $K=20$, no delay ambiguous period can be observed from the likelihood function $\mathcal{L}(\tau)$. By comparison, the UDR is 1~ns for UFS and 5~ns for CFS and NFS given $K=20$. This verifies our delay-ambiguity-free design. Also, we notice that when the number of frequency samples $K$ decreases, the power of sidelobes floor will rise and become comparable to the mainlobe, which may cause the estimated multipath parameters inaccurate when the inference is not completely canceled in the E step. Therefore, through our proposed PFS has no delay ambiguity, which means any small $K$ is theoretically adoptable in the channel sounding, a larger $K$ is still beneficial for multipath extraction.

 \par
\begin{figure}[!ht]
\centering
    \includegraphics[width=0.45\textwidth]{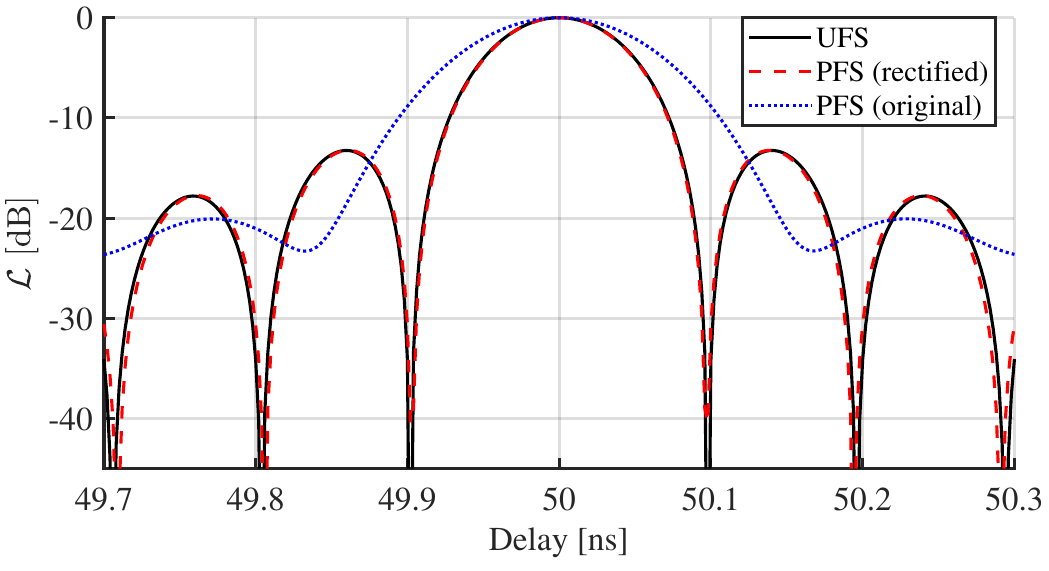}%

    \caption{Zoomed-in mainlobe of likelihood function $\mathcal{L}(\tau)$ for a path delay of 50~ns with $K=50$ under (a) Uniform frequency sampling not affected by molecular absorption, (b) Parabolic frequency sampling with molecular absorption rectified by $I(f;\tau)$ and (c) Parabolic frequency sampling with molecular absorption and without rectification. ($f_\text{c}=375~\text{GHz}$, $B=10~\text{GHz}$, $T=30^\circ \mathrm{C}$ and $D_\text{VP}=20~\text{g/m}^3$)}
    \label{fig:L-compare}
\end{figure}

 Figure~\ref{fig:L-compare} shows the comparison of the zoomed-in single-path likelihood function under uniform frequency sampling, and parabolic frequency sampling. By comparing Fig.~\ref{fig:L-compare}(a), (b) and (c), we observe that the mainlobe of the likelihood function of PFS rectified by factor $I(f;\tau)$ is sharp and similar to that of UFS. Without rectification, the mainlobe of the likelihood function of PFS is 1.5 times wider. This demonstrates that the proposed post-processing factor $I(f;\tau)$ can maintain the mainlobe shape of the likelihood function as uniform frequency sampling and eliminate the effect of molecular absorption.

\subsection{Multipath Extraction Performance}
To verify the proposed PFS and LR-SAGE algorithm, we first use synthesized multipath channel to test the multipath extraction results for different frequency sampling schemes and SAGE algorithms. In the simulation, the multipath channel parameters and the RF parameters are summarized in Table~I. Also, for calculating molecular absorption, the temperature $T=30^\circ \mathrm{C}$ and the water vapor density $D_\text{VP}=20~\text{g/m}^3$. In this work, the performance is evaluated using the root-mean-square error (RMSE) of the estimated path delays. The RMSE is computed for the five most dominant paths from the SAGE algorithm, which are selected by sorting the estimated path amplitudes in descending order. The error is calculated relative to the true path delays specified in Table~\ref{tab:simulation_parameters}.
\begin{table}[!ht]
\renewcommand{\arraystretch}{1.3}
\caption{Key Simulation Parameters for SAGE Algorithm}
\label{tab:simulation_parameters}
\centering

\begin{tabular}{l c}
\toprule[1.5pt] 
\textbf{Parameter} & \textbf{Value} \\
\midrule
\multicolumn{2}{c}{\textbf{Multipath Channel Parameters}} \\
Number of Paths ($L$) & 5 \\
Path Delays ($\tau_l$) [ns] & \{5, 50, 100, 150, 200\} \\
Path Amplitudes ($|\alpha_l|$) & \{1, 0.3, 0.1, 0.07, 0.03\} \\
Path Phases ($\angle\alpha_l$) [rad] & \{0, $\pi/4$, $\pi/4$, $-\pi/3$, $-\pi/3$\} \\
\midrule
\multicolumn{2}{c}{\textbf{RF and System Parameters}} \\
Center Frequency ($f_\text{c}$) & 380 GHz \\
Total Bandwidth ($B$) & 20 GHz \\
Signal-to-Noise Ratio (SNR) & 50 dB \\
\bottomrule[1.5pt] 
\end{tabular}
\end{table}

\par
\begin{figure}[!ht]
\centering
        \includegraphics[width=0.4\textwidth]{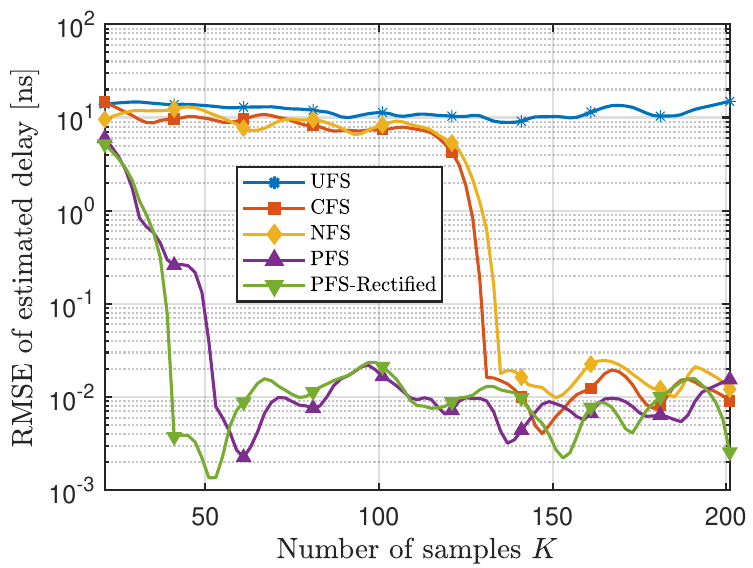}%
    \caption{RMSE of estimated delay $\hat\tau$ by SAGE algorithm versus the number of samples $K$ without molecular absorption.}
    \label{fig:SAGE-NOMA}
\end{figure}

 Figure~\ref{fig:SAGE-NOMA} shows delay RMSE performance for different frequency sampling schemes and SAGE algorithms when the molecular absorption is omitted. The results of 'UFS', 'CFS', 'NFS', 'PFS' are obtained via original SAGE algorithm while the result of 'PFS-Rectified' is from LR-SAGE algorithm. The key observations are as follows: first, UFS which has the shortest UDR cannot correctly estimate the MPCs as the number of frequency samples in the simulation are too small compared with required $K\geq 4000$. Second, the RMSE of CFS and NFS starts to drop rapidly when $K$ exceeds 127 which is the required minimum number of frequency samples for the two schemes. Third, the proposed PFS shows the best delay estimation performance among all the frequency sampling schemes as its RMSE rapidly decreases with $K\geq40$ and shows the lowest RMSE value. Furthermore, it is observed that the LR-SAGE algorithm that rectifies the likelihood function shows slight advantage in the absence of molecular absorption compared with the original SAGE algorithm as it is capable of dealing with the distortion. 

\begin{figure}[!ht]
\centering
        \includegraphics[width=0.38\textwidth]{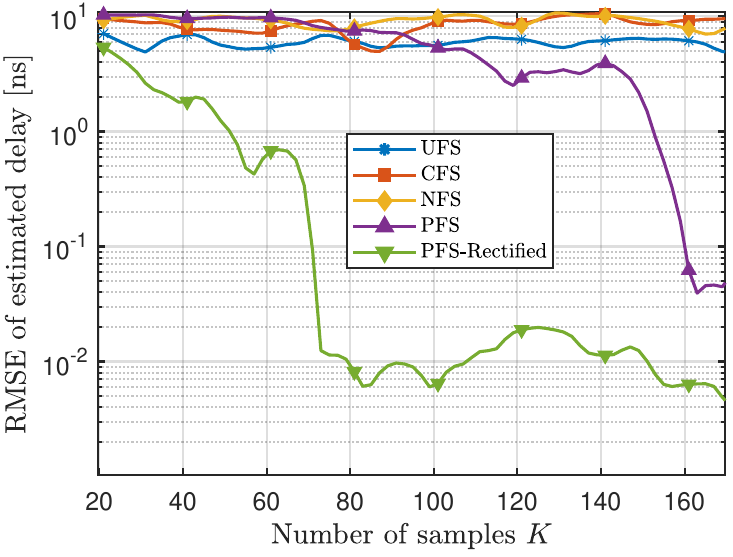}%
    \caption{RMSE of estimated delay $\hat\tau$ by SAGE algorithm versus the number of samples $K$ with molecular absorption ($T=30^\circ \mathrm{C}$ and $D_\text{VP}=20~\text{g/m}^3$).}
    \label{fig:SAGE-MA}
\end{figure}
In Fig.~\ref{fig:SAGE-MA}, we plot the delay RMSE performance considering the molecular absorption. The results show all the frequency sampling schemes, i.e., UFS, CFS, NFS and PFS, using original SAGE algorithm fail to correctly estimate the MPCs. This confirms that molecular absorption, with its frequency dependent attenuation, significantly hinders accurate multipath extraction. Fortunately, our proposed LR-SAGE algorithm is robust to the molecular absorption effect and its delay RMSE is below $10^{-2}$ ns when $K\geq 70$. We also note that due to the molecular absorption, the LR-SAGE algorithm need a slightly larger number of frequency samples to reach a low delay RMSE compared with the case without molecular absorption, as the interference cancellation in the E-step is not perfectly performed in practice.

\begin{figure}[!ht]
\centering
\includegraphics[width=0.48\textwidth]{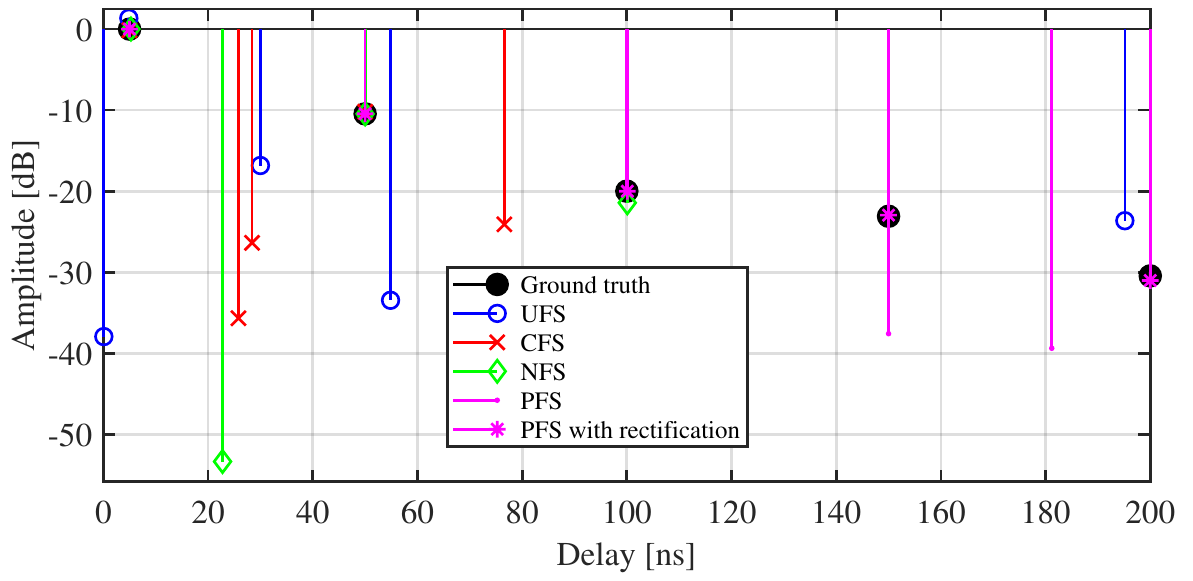}%
    \caption{Multipath extraction results from SAGE algorithm with molecular absorption and $K=100$.}
    \label{fig:SAGE-cir}
\end{figure}

We present an example of MPCs extraction results with molecular absorption and $K=100$ in Fig.~\ref{fig:SAGE-cir}. It shows that only the proposed PFS along with the LR-SAGE algorithm performs an accurate multipath estimation. The other sampling schemes with the original SAGE algorithm deliver inaccurate estimated delay and amplitude of MPCs. This result emphasizes the necessity of the proposed LR-SAGE algorithm in the condition of molecular absorption effect and nonuniform frequency sampling.

\section{Experimental Verification}
In this section, we establish VNA-based THz ISAC channel sounder and conduct channel measurement to validate the proposed sparse frequency sampling scheme and LR-SAGE algorithm.
\begin{figure}[!ht]
\centering
\includegraphics[width=0.4\textwidth]{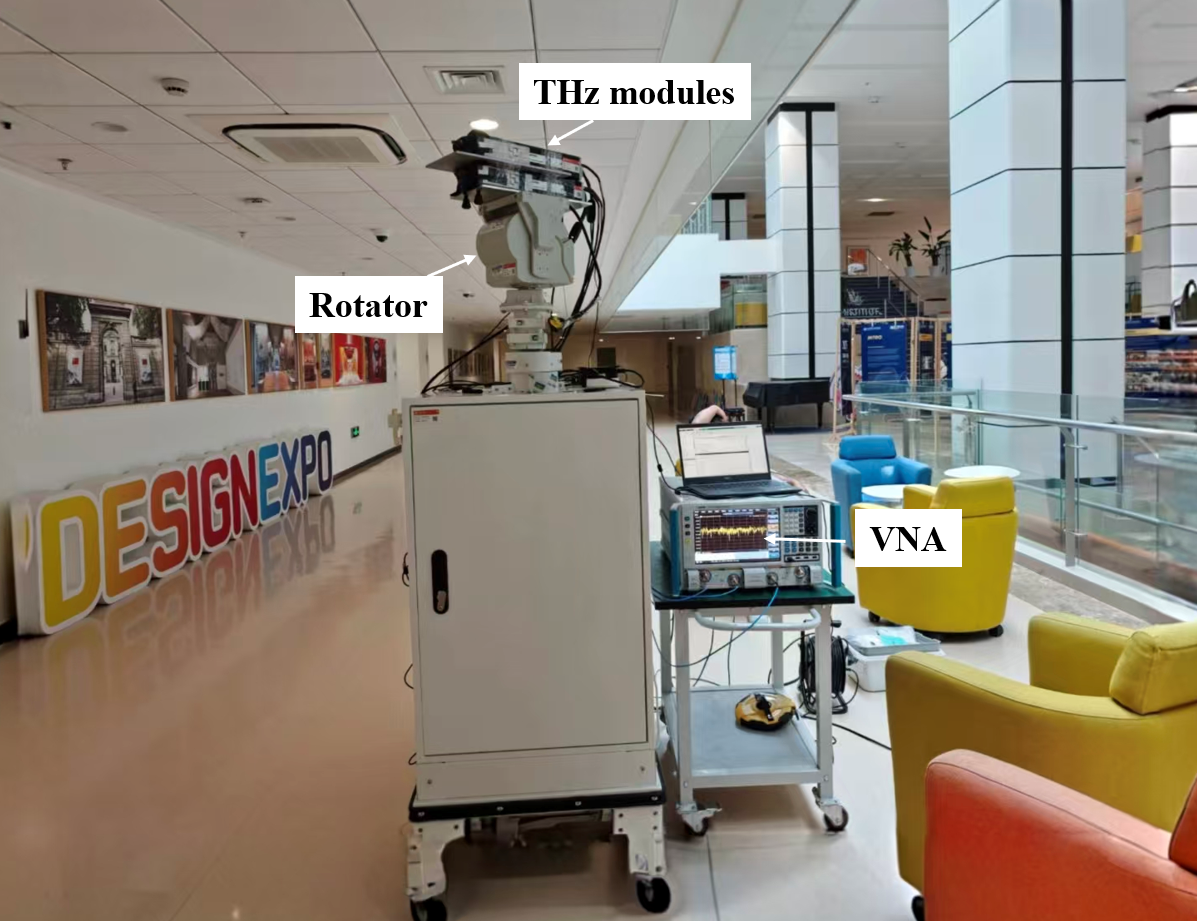}%
    \caption{THz ISAC channel sounder at 280-300~GHz based on VNA.}
    \label{fig:sounder}
\end{figure}
\subsection{Channel Sounder and Scenario}
To validate proposed sparse frequency sampling design with real-world experiment channel data, we establish a VNA-based THz channel sounder at 280-300~GHz. The channel sounder is designed for mono-static channel sounding and consists of a VNA (Ceyear 3672C), THz extension TRx modules, and a rotator to scan azimuth angles as shown in Fig.~\ref{fig:sounder}. The channel measurement is conducted on a floor of an office building in Shanghai Jiao Tong University, Shanghai, China. In the measurement, the rotator enables a $360^\circ$ angle scanning in the azimuth angle domain with a step of $5^\circ$. To validate the proposed PFS along with the LR-SAGE algorithm, we conduct the channel measurement for one specific location at 280-300~GHz in two configurations, respectively. The first configuration is the uniform frequency sampling with $K=12001$ which guarantees a maximum delay of 600~ns. The second configuration is the proposed parabolic frequency sampling with $K=251$, which is conducted under the assistance of technical support from the VNA vendor.

\subsection{Measurement Results}
\begin{figure}[!t] 
\centering
\includegraphics[width=0.45\textwidth]{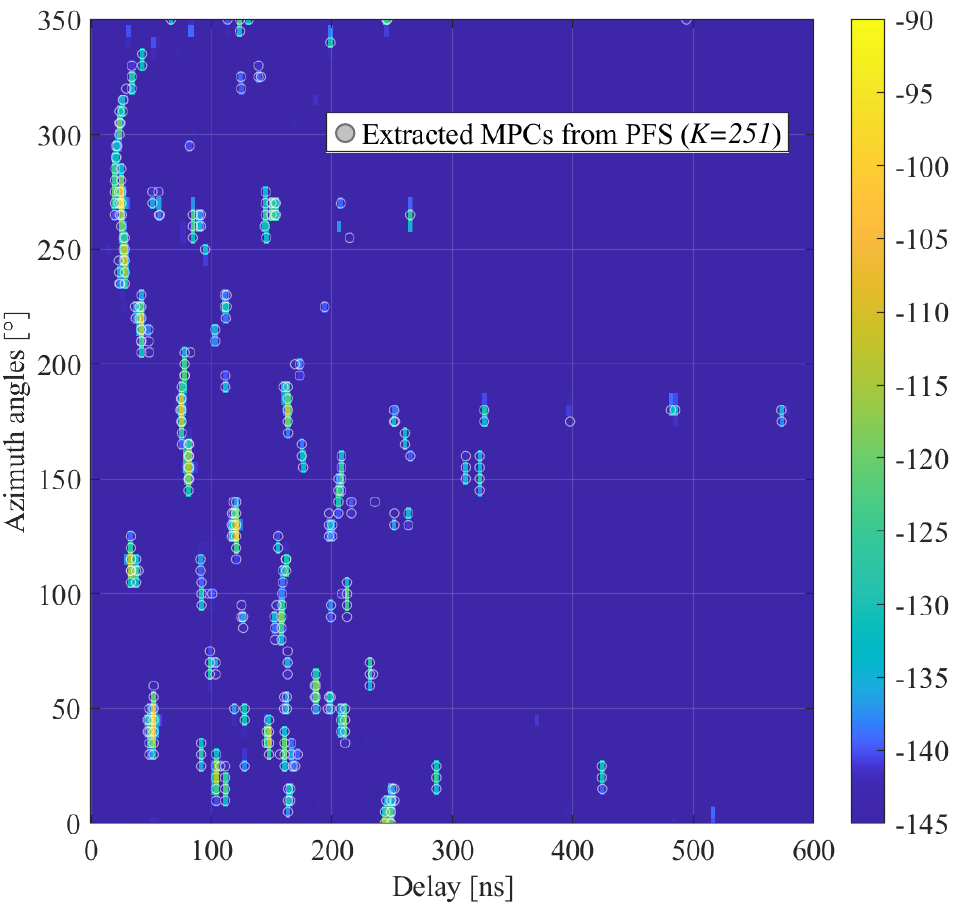}%
    \caption{Measured power-delay-angular profile from UFS with $K=12001$ and extracted MPCs from PFS with $K=251$ and LR-SAGE algorithm.}
    \label{fig:SAGE-pdap}
\end{figure}

In Fig.~\ref{fig:SAGE-pdap} we show the measured power-delay-angular profile (PDAP) from uniform frequency sampling with 12001 frequency samples in each azimuth angle. A large number of MPCs due to multi-reflection and scattering are obvious in the PDAP in the measured 280-300~GHz. Meanwhile, we conduct the LR-SAGE algorithm to extract MPCs from the proposed PFS with 251 frequency samples, denoted by white circles in the figure. The MPCs extracted using only 251 samples align remarkably well with the high-resolution PDAP derived from 12001 samples, which verifies the proposed PFS and the LR-SAGE algorithm. In addition, as the computational complexity of the SAGE algorithm scales with the square of the number of frequency samples, the proposed PFS leads to a 99.96\% reduction in the computational complexity of the LR-SAGE algorithm relative to conventional UFS.

\begin{figure}[!ht]
\centering
\includegraphics[width=0.44\textwidth]{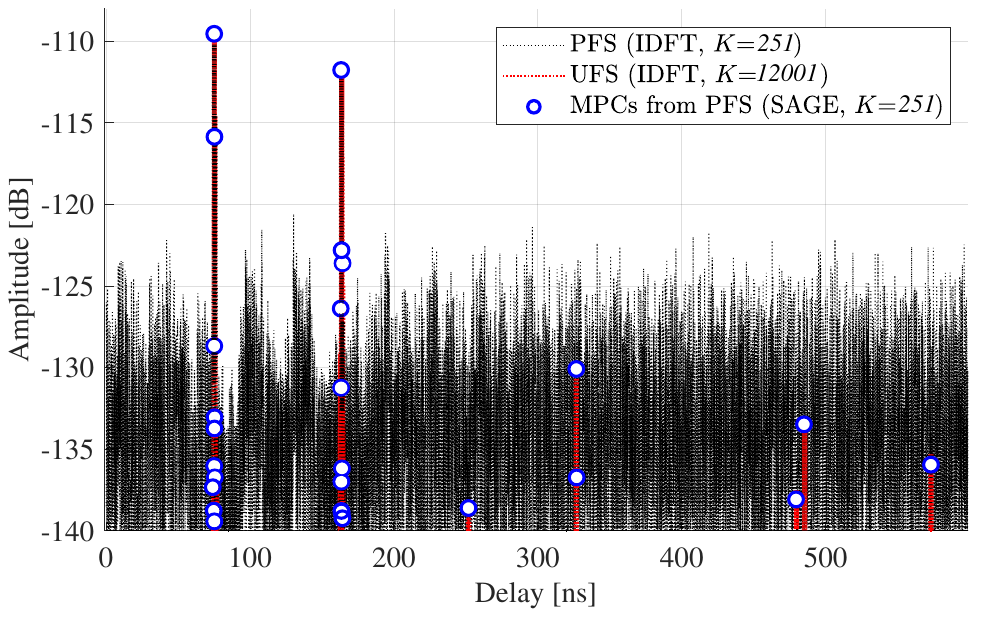}%
    \caption{Measured channel impulse responses of PFS ($K=251$) and UFS ($K=12001$) via IDFT along with the extracted MPCs from PFS and LR-SAGE algorithm.}
    \label{fig:SAGE-mcir}
\end{figure}
Next, in Fig.~\ref{fig:SAGE-mcir} we plot channel impulse responses (CIRs) for one azimuth angle measured from PFS and UFS by the inverse discrete Fourier transform (IDFT). First, in this azimuth angle, 6 dominant MPCs clusters are observed from the CIR of UFS with delay ranging from 80~ns to 580~ns, and amplitude ranging from -110~dB to-140~dB. This shows that back scattering is significant in such high frequency of 280-300~GHz. The two strongest MPCs are closely followed by some weak MPCs, which is caused by the harmonics and diffuse scattering as explained in~\cite{pch}. Second, the CIR obtained from the proposed sparse PFS shows a high noise level compared with that of UFS. This is due to the fact that the sidelobe level of the single-path likelihood of proposed PFS is 20~dB lower than mainlobe when $K=251$. Thus, the whole CIR is corrupted by the sidelobes of the two strongest MPCs and it is impossible to extract effective MPCs directly from the CIR of the sparse PFS. Third, the extracted MPCs from PFS with 251 frequency samples by the LR-SAGE algorithm are also plotted in the figure. The extracted MPCs achieve a dynamic range exceeding 30~dB and are in good agreement with the measured CIR. This occurs because the SAGE algorithm decomposes the parameter estimation of all the paths into that of each single-path, and thereby, avoids the inference among the MPCs. The measurement results not only validate the proposed PFS and LR-SAGE algorithm, but also demonstrate the necessity of the LR-SAGE algorithm. That is, the proposed sparse nonuniform frequency sampling scheme must be conducted along with the proposed LR-SAGE algorithm.

\begin{figure}[h]
\centering
    \subfloat[Path loss]{%
        \includegraphics[width=0.4\textwidth]{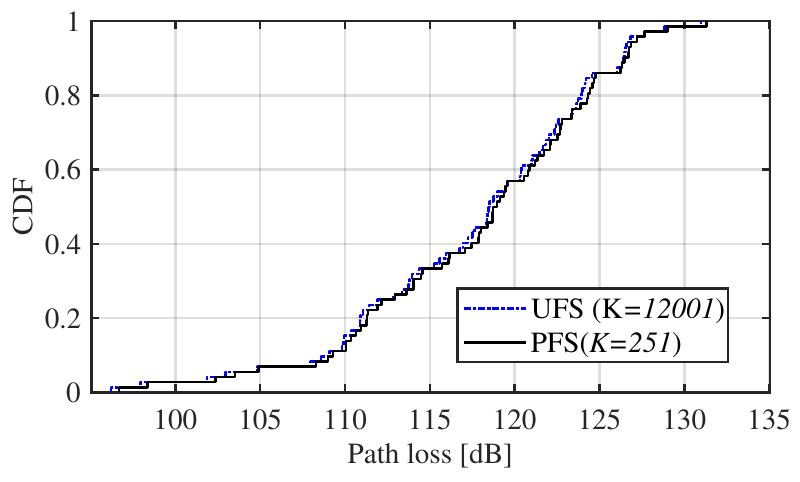}%
    }
    
    \subfloat[Delay spread]{%
\includegraphics[width=0.4\textwidth]{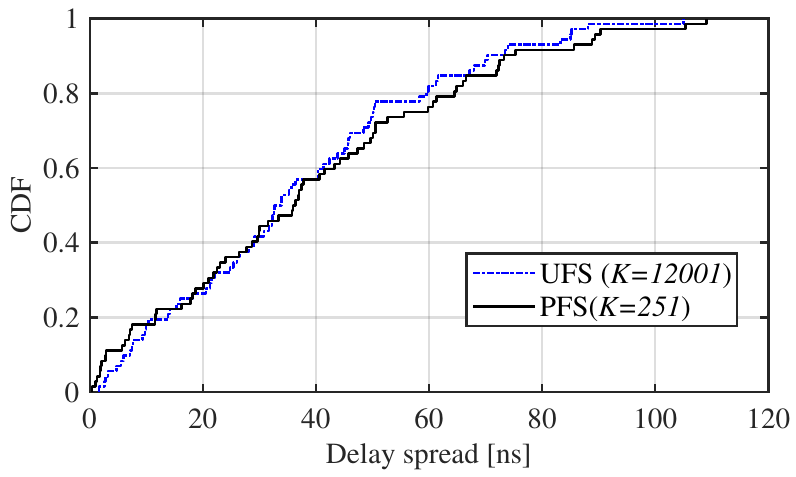}%
    }
    \caption{Channel characteristics comparison of path loss and delay spread between dense UFS and proposed sparse PFS based on channel measurement data.}
    \label{fig:statistics}
\end{figure}

To further validate the channel characteristics from our proposed PFS and LR-SAGE algorithm, we calculate the cumulative density function (CDF) of two critical channel characteristics, i.e., path loss and RMS delay spread in Fig.~\ref{fig:statistics}(a) and (b), respectively. To be concrete, the path loss is calculated by summing up the power of all the MPCs. RMS delay spread is calculated as the square root of the second central moment of all the MPCs delays with power weighted. MPCs of `UFS' is obtained by directly IDFT while the MPCs of `PFS' is extracted by LR-SAGE algorithm. By comparing the results from UFS and the proposed PFS, we conclude that the proposed PFS and LR-SAGE algorithm can well reproduce the statistics of channel characteristics with much fewer frequency samples, which guarantees the validity channel model based on the proposed sparse frequency sampling design.

\section{Conclusion}
This paper has introduced a novel sparse sampling framework, comprising a nonuniform frequency sampling strategy and an LR-SAGE multipath extraction algorithm, to enable large-scale long-range channel sounding. To overcome the key limitations of traditional frequency-domain sounding methods, namely delay ambiguity and absorption-induced distortion, we propose a parabolic frequency sampling (PFS) method that eliminates delay ambiguity and
significantly reduces the required number of frequency samples. Complementing this, Complementing this, we develope an LR-SAGE algorithm incorporating a frequency- and delay-dependent weighting scheme to rectify absorption induced distortions in the channel’s likelihood function, achieving robust multipath parameter estimation. Simulation results demonstrated that the proposed PFS approach attains delay estimation accuracy comparable to dense uniform sampling while requiring substantially fewer measurements, effectively circumventing delay ambiguity issues inherent in conventional sparse schemes. Experimental validations conducted in the 280--300~GHz frequency range further confirmed the efficacy of our approach, and reveals that merely 2\% of the full frequency samples sufficed to reconstruct multipath characteristics consistent with exhaustive full-bandwidth measurements. These results illustrate the potential of the proposed methodology for dramatically accelerating channel sounding and enabling efficient collection of large-scale ISAC channel data, thereby facilitating the advancement of future 6G wireless systems. 


\begin{thebibliography}{99}

\bibitem{wcx}
C. -X. Wang et al., ``On the road to 6G: Visions, requirements, key technologies, and testbeds,'' \textit{IEEE Communications Surveys \& Tutorials}, vol. 25, no. 2, pp. 905--974, Feb. 2023.

\bibitem{AI-ISAC}
N. Wu et al., ``AI-enhanced integrated sensing and communications: Advancements, challenges, and prospects,'' \textit{IEEE Communications Magazine}, vol. 62, no. 9, pp. 144--150, Sep. 2024.
\bibitem{critical}
B. Böck, A. Kasibovic, and W. Utschick, ``Wireless channel modeling for machine learning -- A critical view on standardized channel models,'' arXiv preprint arXiv:2510.12279, 2025.

\bibitem{ywf}
W. Yang et al., ``Integrated sensing and communication channel modeling and measurements: Requirements and methodologies toward 6G standardization,'' \textit{IEEE Vehicular Technology Magazine}, vol. 19, no. 2, pp. 22--30, Jun. 2024.


\bibitem{thz-cst}
C. Han et al., ``Terahertz wireless channels: A holistic survey on measurement, modeling, and analysis,'' \textit{IEEE Communications Surveys \& Tutorials}, vol. 24, no. 3, pp. 1670--1707, Jun. 2022.



\bibitem{cxs-rof}
A. W. Mbugua, W. Fan, K. Olesen, X. Cai and G. F. Pedersen, ``Phase-Compensated optical fiber-Based ultrawideband channel sounder,'' \textit{IEEE Transactions on Microwave Theory and Techniques}, vol. 68, no. 2, pp. 636--647, Feb. 2020.

\bibitem{ju-factory}
S. Ju, D. Shakya, H. Poddar, Y. Xing, O. Kanhere and T. S. Rappaport, ``142 GHz sub-terahertz radio propagation measurements and channel characterization in factory buildings,'' \textit{IEEE Transactions on Wireless Communications}, vol. 23, no. 7, pp. 7127--7143, Jul. 2024.

\bibitem{lyj}
Y. Lyu, Z. Yuan, M. Li, A. W. Mbugua, P. Kyösti and W. Fan, ``Enabling long-range large-scale channel sounding at sub-THz bands: Virtual array and radio-over-fiber concepts,'' \textit{IEEE Communications Magazine}, vol. 62, no. 2, pp. 16--22, Feb. 2024.
\bibitem{gk}
K. Guan et al., ``Channel sounding and ray tracing for intra-wagon scenario at mmWave and sub-mmWave bands,'' \textit{IEEE Transactions on Antennas and Propagation}, vol. 69, no. 2, pp. 1007--1019, Feb. 2021.

\bibitem{ju-office}
S. Ju, Y. Xing, O. Kanhere and T. S. Rappaport, ``Millimeter wave and sub-terahertz spatial statistical channel model for an indoor office building,'' \textit{IEEE Journal on Selected Areas in Communications}, vol. 39, no. 6, pp. 1561--1575, Jun. 2021.

\bibitem{zjh}
J. Zhang et al., ``Deterministic ray tracing: A promising approach to THz channel modeling in 6G deployment scenarios,'' \textit{IEEE Communications Magazine}, vol. 62, no. 2, pp. 48--54, Feb. 2024.
\bibitem{na-d2d}
N. A. Abbasi et al., ``THz band channel measurements and statistical modeling for urban D2D environments," \textit{IEEE Transactions on Wireless Communications}, vol. 22, no. 3, pp. 1466--1479, Mar. 2023.
\bibitem{na-cellular}
N. A. Abbasi et al., ``THz band channel measurements and statistical modeling for urban micro cellular environments,'' \textit{IEEE Transactions on Wireless Communications}, vol. 23, no. 7, pp. 6719--6734, Jul. 2024.
\bibitem{cy-200G}
Y. Chen, C. Han, Z. Yu and G. Wang, ``Channel measurement, characterization, and modeling for terahertz indoor communications above 200 GHz,'' \textit{IEEE Transactions on Wireless Communications}, vol. 23, no. 6, pp. 6518-6532, Jun. 2024
\bibitem{na-icc}
N. A. Abbasi et al., ``Double directional channel measurements for THz communications in an urban environment,'' \textit{Proc. of 2020 IEEE International Conference on Communications (ICC)}, Dublin, Ireland, 2020, pp. 1--6.

\bibitem{kurner}
S. Priebe and T. Kurner, ``Stochastic modeling of THz indoor radio channels,'' \textit{IEEE Transactions on Wireless Communications}, vol. 12, no. 9, pp. 4445--4455, Sep. 2013.

\bibitem{molisch}
A. F. Molisch, J. Gomez-Ponce, N. Abbasi, W. Choi, G. Xu and C. J. Zhang, ``Properties of sub-THz propagation channels and their impact on system behavior: channel measurements and transmission experiments,'' \textit{IEEE Wireless Communications}, vol. 31, no. 1, pp. 18--24, Feb. 2024.


\bibitem{multi-ray}
C. Han, A. O. Bicen and I. F. Akyildiz, ``Multi-ray channel modeling and wideband characterization for wireless communications in the terahertz band,'' \textit{IEEE Transactions on Wireless Communications}, vol. 14, no. 5, pp. 2402-2412, May 2015.

\bibitem{bcs}
S. Ji, Y. Xue and L. Carin, ``Bayesian compressive sensing,'' \textit{IEEE Transactions on Signal Processing}, vol. 56, no. 6, pp. 2346--2356, Jun. 2008.
\bibitem{sbi}
Z. Yang, L. Xie and C. Zhang, ``Off-grid direction of arrival estimation using sparse Bayesian inference,'' \textit{IEEE Transactions on Signal Processing}, vol. 61, no. 1, pp. 38--43, Jan. 2013.
\bibitem{unequal}
A. Ishimaru and Y.-S. Chen, ``Thinning and broadbanding antenna arrays by unequal spacings,'' \textit{IEEE Transactions on Antennas and Propagation}, vol. 13, no. 1, pp. 34--42, Jan. 1965.

\bibitem{Superresolution}
H. Yamada, M. Ohmiya, Y. Ogawa and K. Itoh, ``Superresolution techniques for time-domain measurements with a network analyzer,'' \textit{IEEE Transactions on Antennas and Propagation}, vol. 39, no. 2, pp. 177-183, Feb. 1991.

\bibitem{cs-sfcw}
A. Bayu Suksmono, E. Bharata, A. Andaya Lestari, A. G. Yarovoy, and L. P. Ligthart, ``Compressive stepped-frequency continuous-wave ground-penetrating radar,'' \textit{IEEE Geosci. Remote Sens. Lett.}, vol. 7, no. 4, pp. 665–-669, Oct. 2010.

\bibitem{cs-thz}
S. Hu, C. Shu, Y. Alfadhl and X. Chen, ``Advanced THz MIMO sparse imaging scheme using multipass synthetic aperture focusing and low-rank matrix completion techniques,'' \textit{IEEE Transactions on Microwave Theory and Techniques}, vol. 70, no. 1, pp. 659--669, Jan. 2022.
\bibitem{nested}
P. Pal and P. P. Vaidyanathan, ``Nested arrays: A novel approach to array processing with enhanced degrees of freedom,'' \textit{IEEE Transactions on Signal Processing}, vol. 58, no. 8, pp. 4167-–4181, Aug. 2010.
\bibitem{coprime}
P. P. Vaidyanathan and P. Pal, ``Sparse sensing with co-prime samplers and arrays,’’ \textit{IEEE Transactions on Signal Processing}, vol. 59, no. 2, pp. 573–-586, Feb. 2011.
\bibitem{coarray}
M. Wang and A. Nehorai, ``Coarrays, MUSIC, and the Cramér–Rao bound,'' \textit{IEEE Transactions on Signal Processing}, vol. 65, no. 4, pp. 933--946, Feb. 2017.
\bibitem{bcoprime}
Q. Liu, N. Chu, L. Yu, Z. Shao, H. Qin, and P. Wu, ``A Bayesian framework of non-synchronous measurements at coprime positions for sound source localization with high resolution,'' \textit{IEEE Transactions on Instrumentation and Measurement}, vol. 72, pp. 1–-17, Nov. 2023.

\bibitem{radar-coprime}
H. Pan, J. Pan, X. Zhang and Y. Wang, ``Time-delay estimation of ground-penetrating radar using co-prime sampling strategy via atomic norm minimization,'' \textit{IEEE Transactions on Instrumentation and Measurement}, vol. 73, pp. 1--12, Mar. 2024.
\bibitem{PSF}
A. Ishimaru, ``Unequally spaced arrays based on the Poisson sum formula,'' \textit{IEEE Transactions on Antennas and Propagation}, vol. 62, no. 4, pp. 1549--1554, April 2014.
\bibitem{SFCW-PSF}
F. Gumbmann and A. Schiessl, ``Short-range imaging system with a nonuniform SFCW approach,'' \textit{IEEE Transactions on Microwave Theory and Techniques}, vol. 65, no. 4, pp. 1345--1354, Apr. 2017.
\bibitem{SFCW-SQRT}
W. Si, X. Zhuge and J. Miao, ``Sparse nonuniform frequency sampling for fast SFCW radar imaging,'' \textit{IEEE Access}, vol. 8, pp. 126573--126581, Jul. 2020.
\bibitem{SAGE}
X. Yin, L. Ouyang and H. Wang, ``Performance comparison of SAGE and MUSIC for channel estimation in direction-scan measurements,'' \textit{IEEE Access}, vol. 4, pp. 1163--1174, Mar. 2016.
\bibitem{DSS-SAGE}
Y. Li, et al, ``DSS-o-SAGE: Direction-scan sounding-oriented SAGE algorithm for channel parameter estimation in mmWave and THz bands,'' \textit{IEEE Transactions on Antennas and Propagation}, vol. 73, no. 4, pp. 1969--1983, Apri. 2025.

\bibitem{pch}
T. Zhang et al., "Indoor channel measurements and characterization for virtual multi-antenna at 260-400 GHz," \textit{IEEE Antennas and Wireless Propagation Letters}, to appear.
\end{thebibliography}
\end{document}